\pdfoutput=1
\documentclass[
a4paper,
envcountsame
,final ]{llncs}

\usepackage[style=lncs,
]{biblatex}
\usepackage{doi}

\usepackage{ifdraft}
\newbool{doubleblind}
\newbool{arxiv}

\doubleblindfalse
\arxivtrue

\ifbool{arxiv}
{
  \usepackage[notextcomp]{stix}

\usepackage[scaled=.95]{sourcesanspro}
  \usepackage[scr=boondoxo]{mathalpha}
  \newcommand{\oc}{{!}}
  
  \newcommand{\cons}{\mathbin{{:}\mkern-4mu{:}}}
}
{
  \usepackage{lmodern}
  \usepackage{cmll}
  \usepackage{mathrsfs}
  \newcommand{\eqdef}{\overset{\smash{\scriptscriptstyle\mathrm{def}}}{=}}
  \newcommand{\cons}{\mathbin{{:}{:}}}
  \usepackage{amssymb}
}{}

\usepackage{hyperref}
\usepackage{xcolor}
\definecolor{dark-red}{rgb}{0.4,0,0}
\definecolor{dark-blue}{rgb}{0,0.0,0.35}
\definecolor{blue}{rgb}{0,0,0.7}
\hypersetup{ocgcolorlinks,pdfusetitle,
  bookmarks=true,bookmarksnumbered=false,bookmarksopen=false,
  breaklinks=true,pdfborder={0 0 0},pdfborderstyle={},backref=page,
  colorlinks=true,linkcolor={dark-red},citecolor={dark-blue},urlcolor={blue}}

\ifbool{doubleblind}{
  \usepackage{lineno}
  \linenumbers
}{}

\newcommand{\mypar}[1]{\subsection{#1}}

\usepackage{bussproofs}
\usepackage{amsmath}
\usepackage{mathtools}
\usepackage{stmaryrd}
\usepackage{csquotes}
\usepackage[english]{babel}
\usepackage{mdframed}
\newmdenv[shadow=true,shadowsize=2pt,shadowcolor=black!8
]{framed}

\usepackage{tikz-cd}
\usepackage{listings}
\ifbool{arxiv}{
}{
  \usepackage[textsize=small,obeyFinal]{todonotes}\setlength{\marginparwidth}{8.5em}\setlength{\marginparsep}{2em}
}
\usepackage{cleveref}
\AddToHook{cmd/appendix/before}{\crefalias{section}{appendix}}
\AtBeginEnvironment{appendices}{\crefalias{section}{appendix}}

\ifdraft{\raggedbottom}{}\usepackage{microtype}

\DeclareMathOperator{\CtxComp}{@}
\DeclareMathOperator{\New}{\mathbf{new}}
\DeclareMathOperator{\Delete}{\mathbf{delete}}
\DeclareMathOperator{\NewFail}{Alloc\_failure}
\DeclareMathOperator{\Nil}{\mathrm{Nil}}

\newcommand{\Em}{\mathcal{E}}
\newcommand{\Dm}{\mathcal{D}}
\newcommand{\Ccat}{\mathscr{C}}

\newcommand{\BaseCat}{\mathbf{[R]}}
\usepackage{xspace}
\newcommand{\Move}{\ensuremath{\mathbf{move}}\xspace}
\newcommand{\LinLang}{\ensuremath{\mathrel{\boldsymbol{\mathcal{L}}}}\xspace}
\newcommand{\OLinLang}{\ensuremath{\mathrel{\boldsymbol{\mathcal{O}}}}\xspace}
\newcommand{\AffLang}{\ensuremath{\mathrel{\boldsymbol{\mathcal{O}}_{\boldsymbol{\Em,\Move}}}}\xspace}
\newcommand{\OAffLang}{\ensuremath{\mathrel{\boldsymbol{\mathcal{O}}_{\boldsymbol{\Em}}}}\xspace}
\newcommand{\Set}{\mathcal{S}\mkern-1mu\mathit{et}}

\newcommand{\vdashp}{\vdash_p}
\newcommand{\vdasho}{\vdash_o}
\newcommand{\vdashl}{\vdash_l}

\newcommand{\Up}{\mathord{\Uparrow}}
\newcommand{\Down}{\mathord{\Downarrow}}
\newcommand{\up}{\mathord{\uparrow}}
\newcommand{\down}{\mathord{\downarrow}}

\DeclareMathOperator{\Let}{\mathrm{let}}
\DeclareMathOperator{\Try}{\mathbf{try}}
\DeclareMathOperator{\Raise}{\mathrm{raise}}
\DeclareMathOperator{\RaiseB}{\mathbf{raise}}
\DeclareMathOperator{\Swap}{\mathrm{swap}}
\DeclareMathOperator{\Drop}{\mathrm{drop}}
\DeclareMathOperator{\DropB}{\mathbf{drop}}
\DeclareMathOperator{\Unwind}{\mathrm{unwind}}
\DeclareMathOperator{\DropCtx}{\mathrm{drop\_ctx}}

\DeclareMathOperator{\Match}{\mathrm{match}}
\DeclareMathOperator{\Unless}{\mathbf{unless}}
\DeclareMathOperator{\Coerc}{\mathrm{coerc}}

\DeclareMathOperator{\In}{\mathbin{in}}
\newcommand{\With}{\mathbin{\&}}
\newcommand{\Concat}{\mathbin{+\kern -0.4em+}}
\newcommand{\RmapSym}{\multimap}
\newcommand{\LmapSym}{\multimapinv}
\newcommand{\Rmap}[2]{#1 \RmapSym #2}
\newcommand{\Lmap}[2]{#2 \LmapSym #1}

\def\Colon{\mathbin{:}}

\usepackage{perfectcut}
\newcommand{\Paren}[1]{\perfectunary{IncreaseHeight}(){#1}}

\newcommand{\Bigmachine}[4]{\nthleft{#1}\langle #2\mathrel{\nthmiddle{#1}\vert}#3\mathrel{\nthmiddle{#1}\vert}#4\nthright{#1}\rangle}

\newcommand{\vRuleSpace}{\vspace{0.7em}}

\def\letin#1#2{\Let\,#1=#2\,\In\,}
\def\Interp#1{\left\llbracket#1\right\rrbracket}

\ifdraft{
    \usepackage{outlines}

    \usepackage[normalem]{ulem}
    \def \ifempty#1{\def\temp{#1} \ifx\temp\empty }
    \newcommand{\xnote}[3][]{\textcolor{red}{#2}$\rightarrow$ \textcolor{blue}{\textbf{#1} #3}}
    \newcommand{\xtodo}[2][]{\xnote[#1]{}{#2}}

    \newcommand{\xreplace}[3][]{\ifempty {#3} \textcolor{red}{\textbf{#1} \sout{#2}} \else \textcolor{red}{\sout{#2}} $\rightarrow$ \textcolor{blue}{\textbf{#1} "#3"} \fi}
    
    \newcommand{\xadd}[2][]{\xreplace[#1]{}{#2}}

\usepackage{tikzpagenodes}
\usetikzlibrary{calc}
\usepackage{everypage}
\newcommand\drawframe{\begin{tikzpicture}[remember picture,overlay]
    \draw[black, opacity=0.5] ($(current page text area.south west)$) rectangle ($(current page text area.north east)$);
\end{tikzpicture}}
\AddEverypageHook{\drawframe}

} {
  \ifbool{arxiv}{
}{
    \newcommand{\xnote}[3][]{}
    \newcommand{\xtodo}[2][]{}
    \newcommand{\xreplace}[3][]{#2}
    
    \newcommand{\xadd}[2][]{}
    
    \def\1{}
    \def\2{}
    \def\3{}
    \def\4{}
  }
}

\begin{document}

  \author{
    Sidney Congard\inst{1,2}
    \and Guillaume Munch-Maccagnoni\inst{1}
    \and Rémi Douence\inst{2}
  }
  \authorrunning{S. Congard, G. Munch-Maccagnoni and R. Douence}
  \institute{
    INRIA, LS2N CNRS, Nantes, France
    \and IMT Atlantique, Nantes, France}

\title{Linear Effects, Exceptions, and Resource Safety}
\subtitle{A Curry-Howard Correspondence for Destructors\footnote{Slightly longer version with more details of a paper that appeared in ESOP 2026 with the same title (\doi{10.1007/978-3-032-22720-1_8}). To cite this version: \doi{10.48550/arXiv.2510.23517}.}
}
\titlerunning{Linear Effects, Exceptions, and Resource Safety}

\maketitle

\ifbool{arxiv}
{\noindent
  \makebox[\linewidth]{\small January 25, 2026}
}{}

\begin{abstract}

We analyse the problem of combining linearity, effects, and
exceptions, in abstract models of programming languages, as the issue
of providing some kind of strength for a monad $T(- \oplus E)$ in a
linear setting. We consider in particular for $T$ the \emph{allocation
monad}, which we introduce to model and study resource-safety
properties. We apply these results to a series of two linear effectful
calculi for which we establish their resource-safety properties.
The first calculus is a linear (optionally ordered) call-by-push-value
language with two allocation effects $\New$ and $\Delete$. The
resource-safety properties follow from the linear and ordered
character of the typing rules.

\hspace*{1.5em}We then integrate exceptions with linearity and effects by adjoining
default destruction actions to types, as inspired by C++/Rust
destructors. We see destructors as objects $\delta : A\rightarrow TI$
in the slice category over $TI$. This construction gives rise to a
second calculus, the \emph{resource call-by-push-value}, featuring
exceptions and destructors, and whose weakening and exchange rules
perform side-effects. It is therefore affine at the level of types but
ordered at the level of derivations. As in C++ and Rust, a ``move''
operation---the side-effecting exchange rule---is necessary for
releasing resources in random order, as opposed to LIFO order.
\end{abstract}

\section{Introduction}
\label{sec:intro}

The application of monads to study effects in programming
languages~\cite{Moggi89computationallambda-calculus,Moggi91Monad,Filinski94representingmonads,Power1997,Plotkin_2002,Levy99CBPV,Levy04CBPV},
as well as the application of linearity to study resource-sensitive
aspects of
computation~\cite{Gir87,Lafont1988,Baker1992,Filinski92Linear,Maraist94call-by-namecall-by-value,Berdine2000,Hofmann2000},
are well-established. However, the combination of effects and
resources, despite receiving some attention
\cite{Power2002,Hasegawa2002,Hasegawa2004,Selinger2008,Mellies2012,Moegelberg2014,CFM15},
has much less developed theory and case studies.

In order to understand why the combination of effects and resources
poses new challenges, it is useful to remind that a monad $T$
modelling computational effects is given by an endofunctor on a
cartesian category $\Ccat$, together with families of maps
\begin{align}
  \eta_A &: A\rightarrow TA &&\mbox{and} &\mu_A &: TTA\rightarrow TA\label{eq:monad}
\end{align}
natural in $A\in\Ccat$ and satisfying monoid-like laws, and together
with a family of maps called \emph{strength}\vspace*{-1ex}
\begin{align}
  \sigma_{\Gamma,A} : \Gamma\times TA \rightarrow T(\Gamma\times A)\label{eq:strength}
\end{align}
\pagebreak[1]natural in $\Gamma\in\Ccat$ and $A\in\Ccat$ and satisfying four laws
stating the compatibility with $\eta$, $\mu$ and with the monoidal
structure induced by $\times$ and $1$. The general principle is that
two typed expressions of type $\Gamma \vdash A$ and $\Gamma,A \vdash
B$ are allowed to compose as follows:
\begin{align}
  \Gamma & \vdash t:A &&\mbox{and}& \Gamma,x:A & \vdash u:B
  &&\Longrightarrow& \Gamma\vdash \letin{x}{t}u:B\label{eq:composition}
\end{align}
which reflects in their interpretation as morphisms in the Kleisli category of $T$
\begin{align*}
  \Interp t:\Gamma & \rightarrow TA &&\mbox{and}& \Interp u:\Gamma\times A & \rightarrow TB
\end{align*}
through the ability to select $x$ in $u$ where the strength plays an
essential role:
\begin{equation}
  \Interp{\letin{x}{[\;]}u}:\Gamma\times TA \xrightarrow{\sigma_{\Gamma,A}}
  T(\Gamma\times A) \xrightarrow{T\Interp u} TTB\xrightarrow{\mu_B}TB\;.\label{eq:monadic-bind}
\end{equation}

It is also useful to remind that the application of linearity to model
resource-sensitive phenomena of data and computation requires to move
from a cartesian category ($\Ccat,\times,1$) to a symmetric monoidal
category ($\Ccat,\otimes,I$). This modification gives control over
duplication and erasure (corresponding to contraction and weakening in
logic) since the maps\vspace{-1.5ex}
\begin{align}
 & A\rightarrow A\otimes A &&\textrm{and} && A\rightarrow I
\label{eq:dupl-and-erase}
\end{align}
are no longer available for a general $A$. The linear logic viewpoint
nevertheless subsumes the cartesian viewpoint using a resource
modality ``$\oc$''. Such a resource modality is a comonad such that
its Eilenberg-Moore category $\Ccat^\oc$, whose objects are coalgebras
$(A,\tau\Colon A\rightarrow \oc A)$, has a symmetric monoidal
structure that coincides with the symmetric monoidal structure of
$\Ccat$ in terms of the underlying objects, and such that this
symmetric monoidal structure on $\Ccat^\oc$ is
cartesian~\cite{Mellies2003,PAM2009}. Concretely, duplication and
erasure (\ref{eq:dupl-and-erase}) are available whenever $A$ is given
with a map $\tau:A\rightarrow \oc A$ satisfying two laws of compatibility
with the comonad structure. More rarely---but importantly in our
story---the symmetric monoidal structure is sometimes replaced by a
monoidal structure, which amounts to providing control over the
exchange rule in logic since the symmetry
\begin{align}
 & A\otimes B \rightarrow B\otimes A
\label{eq:exchange}
\end{align}
is no longer available unconditionally. The corresponding logics are
called \emph{ordered} instead of linear. But  again, the exchange
rule can be reintroduced selectively with a resource
modality~\cite{Hasegawa2016}. Ordered logic can be useful to model the
order in the release of
resources~\cite{Polakow2001PhD,Polakow2001,Walker2005a}, as we will
see again in this paper.

\mypar{Linear effects}
In this context, it is tempting to define a notion of \emph{linear
computational effect} to be given again by a monad $(T,\eta,\mu)$
(\ref{eq:monad}) on $\Ccat$ and a strength now given as a family of
maps\begin{align}
  \sigma_{\Gamma,A} : \Gamma\otimes TA \rightarrow T(\Gamma\otimes A)\label{eq:strength-lin}
\end{align}
\pagebreak[1]natural in $\Gamma$ and $A$ and satisfying the same four laws. It is
also called a \emph{left strength} when the monoidal structure
$\otimes$ is non-symmetric. An example of such a strong monad is the
linear state monad $S\multimap (-\otimes S)$ for any $S\in\Ccat$ in
any symmetric monoidal closed category $\Ccat$.
This transposition to the linear context of computational effects
is well-behaved and works similarly to (non-linear) monadic
effects~\cite{Power1997,Hasegawa2002}.

Unfortunately, important examples of effects in models of linear logic
do not have a strength~(\ref{eq:strength-lin}) in such a restrictive
sense, such as the control effects modelled by the following monads in
models of linear logic~\cite{Blute1996,Hasegawa2002,Hasegawa2004}:
\begin{itemize}
\item the monad ``$?$'' of \emph{linearly-defined continuations} (but
not \emph{linearly-used}) $?A \eqdef \oc(A\multimap \bot)\multimap \bot$,
whose effect corresponds to call/cc-style control operators;
\item for any object $E$, the exception monad
$\Em \eqdef (-\oplus E)$, which is used to model error types and
exceptions with $\oplus$ the categorical coproduct---the focus of this paper.
\end{itemize}
At this point, it is instructive to go back to the computational
intuition behind the strength with the interpretation of the monadic
binding~(\ref{eq:monadic-bind}, where $\times$ is now replaced by
$\otimes$). We can see that the parameter $\Gamma$ in
(\ref{eq:strength-lin}) represents the context of variables in the
rest of the computation ($\letin{x}{[\;]}u$) at the location where an
effect is performed. But control effects have this particularity that
they can change how many times the rest
is computed, so we cannot guarantee that variables in $\Gamma$ are not
duplicated nor erased.

\mypar{Strength with respect to a resource modality}\label{sec:strength-wrt} In fact, starting
from this intuition we can suggest a relaxed notion of strength,
something like (\ref{eq:strength-lin}) where $\Gamma$ is assumed to
possess a coalgebra structure $\tau\Colon\Gamma\rightarrow \oc\Gamma$,
so that it can be erased. It so
happens~\cite{Blute1996,Hasegawa2002,Hasegawa2004} that the monad
$?$ of linearly-defined continuations and the exception monad $\Em$
each do have a \emph{(left) strength with respect to $U^\oc$}, where
$U^\oc:\Ccat^\oc\rightarrow\Ccat$ is the forgetful functor of the
category of coalgebras,\footnote{The cited authors actually consider
this notion for the Kleisli adjoint resolution of $\oc$, but it can be
defined similarly for any adjoint resolution, and even any strong
monoidal functor $U$.} defined as a family of maps\vspace*{-1.5ex}
\begin{align}
  \sigma_{\Gamma,A} : U^\oc\Gamma\otimes TA \rightarrow
  T(U^\oc\Gamma\otimes A)\label{eq:strength-relative}
\end{align}
natural in $\Gamma\in\Ccat^\oc$ and $A\in\Ccat$ and satisfying again
four laws of compatibility with $\eta$, $\mu$ and with the monoidal
structures of $\Ccat$ and $\Ccat^\oc$.
If we look concretely at the strength with respect to $U^\oc$ of the
exception monad $\Em$, we see that it consists for each
$A\in\Ccat$ and $\Gamma\in\Ccat^\oc$ of two maps
\begin{align}
  &U^\oc\Gamma\otimes A \rightarrow (U^\oc\Gamma\otimes A)\oplus E
  &&\textrm{and}
  && U^\oc\Gamma\otimes E \rightarrow (U^\oc\Gamma\otimes A)\oplus E\label{eq:strength-exception}
\end{align}
corresponding respectively to the \emph{normal case} and the
\emph{exceptional case} describing the propagation of the exception,
whose copairing must be subject to the mentioned naturality and coherence
conditions. The normal case is given by the left inclusion whereas the
exceptional case is obtained from $U^\oc\Gamma\otimes E \rightarrow E$,
which follows from the erasure (\ref{eq:dupl-and-erase}) given the
hypothesis that the monoidal structure on $\Ccat^\oc$ is cartesian. We
are now ready to make an important observation: for the purpose of
deriving a strength (\ref{eq:strength-relative}) we could consider
other resource modalities than $\oc$; for instance the hypothesis that
$\Ccat^\oc$ is semi-cartesian suffices (the monoidal unit of $\Ccat^\oc$
being a terminal object). This amounts to replacing $\oc$ in
(\ref{eq:strength-relative}) with an affine resource modality that
permits erasure
but not duplication in
(\ref{eq:dupl-and-erase}).

Now that we have described a relaxed notion of strength for the monad
of linearly-defined continuations and the exception monad in a linear
context, we have to mention how restrictive a
strength~(\ref{eq:strength-relative}) seems to be in light of the
interpretation of composition~(\ref{eq:composition}). The notion of
strength with respect to $\oc$ amounts to precluding any linear variable
from appearing in the monadic binding~(\ref{eq:monadic-bind}), in
other words linear variables can only appear in $\Gamma$ if it can be ensured that
no control effect is performed. This models a restrictive approach in
which linearity and control effects are exclusive with one another
(see \cref{sec:discussion}).

\mypar{Destructors}
It is now clear that linearity and control effects do not mix well, or
so it seems. Our goal is to show how linearity and exceptions can
actually be mixed in more important situations than it seems, with a
technique that we explain abstractly but that has been discovered from
practical consideration in the context of resource management by the
designers of the C++ programming
language~\cite{KoenigStroustrup90,Stroustrup1993History,Hinnant2002}.
(We are deferring the more historical discussion to
\cref{sec:discussion}.)

The starting point for describing this important discovery is to
assume a linear model of computation as above given by a symmetric
monoidal closed category $\Ccat$ together with a given notion of
linear effect, that is, a given monad $T$ on $\Ccat$ with a strength
(\ref{eq:strength-lin}). As we have explained, this is a notion of
linear model with linear effect which has previously been
studied~\cite{Power1997,Hasegawa2002,Mellies2012}; in particular a
concrete model of computation is given by a \emph{linear
call-by-push-value} (linear CBPV) calculus~\cite{CFM15} that will provide the basis
of a first linear calculus we study.

In this model, we are interested
in integrating exceptions. Assuming $\Ccat$ has finite coproducts and
that tensor products distribute over the coproducts, this means that
we want to combine the monad $T$ and the exception monad $\Em$ into a strong
monad $T\Em$. Remember that as a general principle of the exception
monad, $T\Em$ has a monad structure arising from a distributive law of
monads $\Em T\rightarrow T\Em$.
However we meet again the obstacle that $\Em$ does not have a strength
in general, so we cannot obtain a strength for $T\Em$ by composition.

But it might still be the case that $T\Em$ has some kind of strength.
This amounts to finding for all $A\in\Ccat$ something like
\begin{align}
  \Gamma\otimes T(A\oplus E) \rightarrow T((\Gamma\otimes A)\oplus E)
\label{eq:strength-TE}
\end{align}
for which, given that $T$ is strong, it suffices to find maps
\begin{align*}
  &\Gamma\otimes A \rightarrow T((\Gamma\otimes A)\oplus E)
  &&\textrm{and}
  &&\Gamma\otimes E \rightarrow T((\Gamma\otimes A)\oplus E)
\end{align*}
subject to some conditions. The one on the left-hand side---the normal
return---is obtained in a straightforward manner by inclusion and unit
of $T$. For the one on the right-hand side---the exceptional
return---observe that the difference with
(\ref{eq:strength-exception}) is the presence of $T$ on the right-hand
side. This second type suggests that $\Gamma$ has to be erasable,
but that erasure is allowed to perform effects in $T$. Indeed, observe
that the second map above can be derived if $\Gamma$ is provided with
some map $\Gamma\rightarrow TI$. We call such an erasure map that
performs an effect a \emph{destructor}, by analogy with C++
destructors.

\mypar{The monoidal category of destructors}
\label{sec:monoidal-destructor-cat}
The desire to find a strength for $T\Em$ therefore suggests to
consider for $\Gamma$ in (\ref{eq:strength-TE}) an object with a given
destructor $\delta:\Gamma\rightarrow TI$, that is to say an object in
the slice category $\Ccat_{/TI}$. Recall that it is the category whose
objects are arbitrary pairs $(A\in\Ccat,\delta:A\rightarrow TI)$, and
whose morphisms are morphisms in $\Ccat$ that preserve the second
component $\delta$.
The slice category $\Ccat_{/TI}$ enjoys a series of nice properties,
the most striking one being that it gives rise to a resource
modality on $\Ccat$~\cite{CombetteMunch18}. Indeed:
\begin{itemize}
\item
$\Ccat_{/TI}$ has a monoidal structure arising from the monoid
structure on $TI$:
\[
TI\otimes TI\xrightarrow{\sigma_{TI,I}} T(TI\otimes
I) \xrightarrow{\cong} TTI \xrightarrow{\mu} TI
\]
The monoidal unit is given by $(I,\eta_I:I\rightarrow TI)$ and, for
$(A,\delta_A:A\rightarrow TI)$ and $(B,\delta_B:B\rightarrow TI)$, the
tensor product is given by:
\[
(A\otimes B,\delta_{A\otimes B}:A\otimes
B\xrightarrow{\delta_A\otimes\delta_B} TI\otimes
TI\rightarrow TI)\;.
\]
In words, $A\otimes B$ has a canonical destructor which releases $B$
and $A$, as we will see in the reverse order of allocation.

\item
We also observe that there is a strong monoidal functor
$U:\Ccat_{/TI}\rightarrow\Ccat$, that sends the monoidal structure of
$\Ccat_{/TI}$ to the one of $\Ccat$ (strictly so).

\item
If we assume that $\Ccat$ has finite products that we note
$(\With,\top)$, then $U$ has a right adjoint
$G:\Ccat\rightarrow\Ccat_{/TI}$ sending objects $A\in\Ccat$ to
\begin{align}
(A\With TI, \pi_2:A\With TI\rightarrow TI)\in\Ccat_{/TI}\label{eq:rightadjoint}
\end{align} and
morphisms $f:A\rightarrow B$ to $f\With TI$. In particular if we note
$\Dm \eqdef UG$ the comonad on $\Ccat$ associated to the adjunction,
one has $\Dm=(-\With TI)$ as our resource modality.

\item
The latter adjunction is comonadic: $\Ccat_{/TI}\simeq\Ccat^\Dm$. This can be observed from the fact that
a coalgebra $(A,\tau:A\rightarrow \Dm A)$ for $\Dm$ boils down to an
object $A$ provided with a morphism $A\rightarrow TI$ in $\Ccat$
without any other condition.
\end{itemize}

\noindent This setup was advocated as a starting point to study C++
destructors in~\cite{CombetteMunch18}.

At this stage we find it useful to sum up the assumptions on $\Ccat$:
a distributive symmetric monoidal category with finite products and a
chosen strong monad $T$. In particular \pagebreak[1]any model of linear logic or
intuitionistic linear logic is suitable (where $T$ can be any linear
state monad, for instance). We have just established that in those
models, $\Dm=(-\With TI)$ has the structure of a resource modality
whose Eilenberg-Moore category is (equivalent to) $\Ccat_{/TI}$.

It is a nice exercise to check that this setup indeed gives a strength
for the composite monad $T\Em$ with respect to
$U:\Ccat_{/TI}\rightarrow\Ccat$, more precisely:

\begin{theorem}
Let $E\in\Ccat$, $\Em$ the exception monad $-\oplus E$, and $T\Em$ the
monad obtained from the distributive law of monads $\Em T\rightarrow
T\Em$. Then the monad $T\Em$ underlies a strong monad on the
$\Ccat^+\mkern-4mu$-category $(\Ccat_{/TI},\ast)$, in the terminology
of Melliès~\cite{Mellies2012}, where $\ast$ is the pseudo-action
defined by composition with $-_1\ast-_2\eqdef (U{-_1}) \otimes -_2$.
\end{theorem}

Now, unlike usual resource modalities that add duplication and/or
erasure (\ref{eq:dupl-and-erase}), \emph{$\Ccat_{/TI}$ is in general
not cartesian nor even semi-cartesian. In fact, its monoi\-dal structure
is not even symmetric} (unless $T$ is commutative). Indeed,
$\delta_{A\otimes B}$ and $\delta_{B\otimes A}$ can in general be
distinguished by the order in which the effects of $\delta_A$ and
$\delta_B$ are performed.

Nevertheless, just as linear models with a resource modality ``$\oc$''
give rise to an intuitionistic model by the Girard
translations~\cite{Gir87}, one can build an \emph{ordered
CBPV} model whose (positive) objects are those of
$\Ccat_{/TI}$, as an instance of a general principle generalising the
Girard translation~\cite{Mellies2012,CFM15} applied to the resource
modality $\Dm$. This motivates the study of a second calculus after
the first one, whose (positive) types are supplied with a destructor
and whose effects include those of $T$ and exceptions.

\mypar{Modelling resources with the allocation monad}
In this new context, notice that everything is linear in the sense
that we did not make use of a resource modality ``$\oc$''. We set out to
show how \emph{resource-safety} properties are thus ensured by
construction. \emph{Resource management} in programming aims to ensure
the correct allocation, use, and release of \emph{resources}, which
are transient values denoting the validity of some state (a memory
allocation, a lock, or typically resources from the operating system).
The need for correct resource management in programming further complicates the
problems with mixing linearity and control: indeed, a need to handle
errors arises from the possibility that resource acquisition can fail,
whereas a need for linearity arises from the fact that the resources
must be released in a timely fashion and no more than once.

We introduce the \emph{allocation monad} as a way to model resource
management and state and establish resource-safety properties. The
allocation monad will play the role of $T$ in the slice category
$\Ccat_{/TI}$. We first assume given an atomic type $R$ of resources
and ask for two effectful operations:
\begin{itemize}
\item $\New: \Rmap{I}{R \oplus I}$ acquires a resource, or fails if there are no
    resources available. Later on, it will also be given with type
    $\Rmap{I}{R}$ when the failure can be represented with an
    exception;
\item $\Delete: \Rmap{R}{I}$ releases the given resource without fail.
\end{itemize}
Notice that no operation is given for interacting with resources: we
are only interested in observing resource-safety properties of the
program state arising from the (correct) use of $\New$ and $\Delete$.
Notice also that we represent failure with an explicit use of the
exception monad $\Em = (-\oplus E)$ for $E \eqdef I$ as an error type. But using
an error type explicitly is not a solution to the lack of exceptions
since we can expect similar obstacles to programming to arise from the
lack of general strength for $\Em$ (see discussion in
\cref{sec:discussion}).

To implement those effectful operations, we then define the allocation
monad as the linear state monad on the type $[R]$ of lists of $R$:
\[
    TA \eqdef \Rmap{[R]}{A \otimes [R]}
\]
As an instance of the linear state monad, it is strong. The morphism
$\New: I\rightarrow T(R \oplus I)$ is defined by popping an element
from the list, or returning the error $I$ if the list is empty.
The morphism $\Delete : R\rightarrow TI$ is defined by pushing the resource onto the list; this is guaranteed to never fail.

Observe that an effectful, closed program of type $A$ in this setup
corresponds to a morphism $I\rightarrow (\Rmap{[R]}{A \otimes [R]})$.
Its execution consists of supplying an initial list of available
resources (\emph{free-list}). Now if the type $A$ is purely positive
and does not contain the type $R$, we expect that \emph{a correct
linear program will have as final free-list the same list as given
initially, up to a permutation}. We can understand this as a
resource-safety property: only resources acquired from the initial
free-list have been released, all resources have eventually been
released including in case of error; no resource has been released
twice.

Moreover, assuming now an ordered rather than linear setting, that is
without symmetry (\ref{eq:exchange}), we expect that \emph{the final
free-list is \emph{identical} to the initial free-list}\label{prop:resource-safety}. This reflects
the observation that without symmetry, resources are managed in a
last-in-first-out (LIFO) order. Ordered logics often have two forms of
closures operating on opposite sides $\Lmap{A}{B}$ and $\Rmap{A}{B}$,
but as we will see this ordered resource-safety property is only
true when one does not have $\Rmap{A}{B}$; hence we only have
$\Lmap{A}{B}$ in this paper for the ordered logics (which still
coincides with $\Rmap{A}{B}$ when adding symmetry).

\mypar{Outline and contributions}
We have just explained abstractly how to mix linearity and exceptions,
and we introduced the \emph{allocation monad} modelling an idealised
global ``free-list'' allocator, with the purpose of studying
resource-safety properties.
In \textbf{\Cref{sec:linear-lang}} we define a
\emph{linear CBPV} calculus \LinLang which has an \emph{ordered} fragment
by removing its exchange rule (\ref{eq:exchange}), hence
named \OLinLang. This calculus features allocation effects ($\New$,
$\Delete$), together with a simple type system and a small-step
operational semantics. We establish elementary properties of \LinLang
and \OLinLang relating typing to reduction.
Then, in \textbf{\Cref{sec:resource-props}} we define and
prove \emph{resource-safety properties} based on the allocation monad.

In \textbf{\Cref{sec:affine-lang}} we present the \emph{resource
CBPV}, an extension of the ordered language, through a
calculus \AffLang. It is an affine CBPV language with
allocation effects and exceptions, in which every positive type has a
chosen effectful erasure map ($\DropB$), and in which allocation
failure results in an exception. Its exchange rule does a side-effect,
as it changes the observable execution order of destructors; we also
consider its ordered fragment \OAffLang. We define the semantics
of \AffLang by a translation into \LinLang that preserves the ordered
fragment.\vspace*{-1.5ex}
\begin{center}
\begin{tikzcd}[column sep = 4em, row sep = 0.75em]
\OAffLang \ar[r,"\llbracket - \rrbracket"] \ar[d,phantom,"\subset"{rotate=270}] & \OLinLang \ar[d,phantom,"\subset"{rotate=270}] \\
\AffLang  \ar[r,"\llbracket - \rrbracket"] & \LinLang
\end{tikzcd}
\end{center}
\AffLang and \OAffLang inherit the resource-safety properties of
\LinLang and \OLinLang, respectively.

Then, in \textbf{\Cref{sec:cat-model}} we define a presheaf model
for \OLinLang,
which models an ordered logic with a destructor for all types, without
being a model of affine logic in a traditional sense.
Lastly, in \textbf{\Cref{sec:discussion}}, we conclude with a
discussion placing the results in the context of the theory of
programming languages.

\section{An ordered effectful calculus of allocations}
\label{sec:linear-lang}

In this section, we describe a calculus with allocation effects
$\New,\Delete$, and with an ordered type system (\OLinLang) which
optionally can be upgraded to linear (\LinLang).
We provide a common small-step operational semantics modelling this
concrete effect. Our starting point is a simplified version of the
linear CBPV calculus~\cite{CFM15}: for simplicity we
consider a calculus given in natural deduction, as opposed to sequent
calculus, as we do not investigate the reductions and conversions on
open terms. Starting from the linear CBPV calculus, this
amounts to considering the derivation of natural deduction rules in
the sequent calculus of \cite{CFM15}, which directly gives an
operational semantics in the form of typed abstract machines in head
reduction. It is then adapted to the allocation effect by adding a
list of resources, the free-list, to every machine configuration.

Ordered logic refines linear logic by removing the exchange rule, due
to which the order of formulae in the antecedent matters for
provability. As in Walker~\cite{Walker2005a}, we consider a sequencing
(composition) rule restricted as follows:
\begin{center}
\AxiomC{$\Gamma, A \vdash B$}
\AxiomC{$\Delta \vdash A$}
\RightLabel{\scriptsize{let}}
\BinaryInfC{$\Gamma, \Delta \vdash B$}
\DisplayProof
\end{center}
Unlike \cite{Polakow2001,Walker2005a}, we consider left functions:
\begin{center}
\AxiomC{$A, \Gamma \vdash B$}
\RightLabel{\scriptsize{$\LmapSym_i$}}
\UnaryInfC{$\Gamma \vdash \Lmap{A}{B}$}
\DisplayProof
\hfil
\AxiomC{$\Gamma \vdash A$}
\AxiomC{$\Delta \vdash \Lmap{A}{B}$}
\RightLabel{\scriptsize{$\LmapSym_e$}}
\BinaryInfC{$\Gamma, \Delta \vdash B$}
\DisplayProof
\end{center}
instead of right functions whose rules are symmetric\ifbool{arxiv}{:
\begin{center}
\AxiomC{$\Gamma, A \vdash B$}
\RightLabel{\scriptsize{$\RmapSym_i$}}
\UnaryInfC{$\Gamma \vdash \Rmap{A}{B}$}
\DisplayProof
\hfil
\AxiomC{$\Gamma \vdash \Rmap{A}{B}$}
\AxiomC{$\Delta \vdash A$}
\RightLabel{\scriptsize{$\RmapSym_e$}}
\BinaryInfC{$\Gamma, \Delta \vdash B$}
\DisplayProof
\end{center}
}{. }The notation for left functions might seem unusual but one can think
of the notation for an exponential object $B^A$, and it is justified
by the isomorphism\ifbool{arxiv}{:
\begin{align*}
\Lmap{A}{(\Lmap{B}{C})}&\cong \Lmap{(B\otimes A)}{C}
& \Rmap{A}{(\Rmap{B}{C})}&\cong \Rmap{(A\otimes B)}{C}
\end{align*}
}{ $\Lmap{A}{(\Lmap{B}{C})}\cong \Lmap{(B\otimes A)}{C}$. }
We will justify the choice of left vs. right functions
in \cref{sec:resource-props}.

CBPV
will play an important role for the translation
in \cref{sec:affine-lang}, and is also interesting in its own right
since its treatment of the arrow type is useful to comprehend the
multiplicity of closure types in the present of first-class resources
(as we will discuss in \cref{sec:discussion}).
Concretely, connectives and types are classified into two kinds,
positive and negative, which require different evaluation strategies.
Notably, the type $R$ of allocated resources is positive.
\[
\text{Types }A,B:\begin{cases}
\text{Positive types } & P,Q \Coloneq R\mid1\mid A\otimes B\mid A\oplus B\\
\text{Negative types } & N,M \Coloneq \Lmap{A}{B} \mid A \With B\\
\end{cases}
\]
Expressions of
positive type evaluate eagerly, whereas expressions of negative type
evaluate lazily. We introduce the notation $\varepsilon \Coloneq
+ \mid -$ for polarities\ifbool{arxiv}{, and associate to each type
its polarity $\varpi(A)$ with $\varpi(P) = +$ and $\varpi(N) = -$}{}.

\begin{figure}[t]
\begin{framed}
\centering

\AxiomC{}
\RightLabel{\scriptsize{var}}
\UnaryInfC{$x:A \vdash x:A$}
\DisplayProof
\hfil
\AxiomC{$\Gamma \vdash t:A$}
\AxiomC{$\sigma \in \Sigma(\Gamma, \Gamma')$}
\RightLabel{\scriptsize{struct}}
\BinaryInfC{$\Gamma' \vdash t[\sigma]:A$}
\DisplayProof
\vRuleSpace

\AxiomC{}
\RightLabel{\scriptsize{$\New$}}
\UnaryInfC{$\vdash \New: \Lmap{1}{(R \oplus 1)}$}
\DisplayProof
\hfil
\AxiomC{}
\RightLabel{\scriptsize{$\Delete$}}
\UnaryInfC{$\vdash \Delete: \Lmap{R}{1}$}
\DisplayProof
\vRuleSpace

\AxiomC{$\Delta \vdash t:A_\varepsilon$}
\AxiomC{$\Gamma, x:A \vdash u:B_{\varepsilon'}$}
\RightLabel{\scriptsize{let}}
\BinaryInfC{$\Gamma, \Delta \vdash (\Let x^\varepsilon=t \In u)^{\varepsilon'}:B$}
\DisplayProof
\vRuleSpace

\AxiomC{}
\RightLabel{\scriptsize{$1_i$}}
\UnaryInfC{$\vdash (): 1$}
\DisplayProof
\hfil
\AxiomC{$\Delta \vdash v:1$}
\AxiomC{$\Gamma, \Gamma' \vdash t:A_\varepsilon$}
\RightLabel{\scriptsize{$1_e$}}
\BinaryInfC{$\Gamma, \Delta, \Gamma' \vdash \delta (v, ().t)^\varepsilon:A$}
\DisplayProof
\vRuleSpace

\AxiomC{$\Gamma \vdash v:A$}
\AxiomC{$\Delta \vdash w:B$}
\RightLabel{\scriptsize{$ \otimes _i$}}
\BinaryInfC{$\Gamma, \Delta \vdash (v, w): A \otimes B$}
\DisplayProof
\hfil
\AxiomC{$\Gamma \vdash v:A$}
\RightLabel{\scriptsize{$ \oplus _{i1}$}}
\UnaryInfC{$\Gamma \vdash \iota_1v: A \oplus B$}
\DisplayProof
\hfil
\AxiomC{$\Gamma \vdash v:B$}
\RightLabel{\scriptsize{$ \oplus _{i2}$}}
\UnaryInfC{$\Gamma \vdash \iota_2v: A \oplus B$}
\DisplayProof
\vRuleSpace

\AxiomC{$\Delta \vdash v:A \otimes B$}
\AxiomC{$\Gamma, x:A, y:B, \Gamma' \vdash t:C_\varepsilon$}
\RightLabel{\scriptsize{$ \otimes _e$}}
\BinaryInfC{$\Gamma, \Delta, \Gamma' \vdash \delta (v, (x,y).t)^\varepsilon:C$}
\DisplayProof
\vRuleSpace

\AxiomC{$\Delta \vdash v:A \oplus B$}
\AxiomC{$\Gamma, x:A, \Gamma' \vdash t:C_\varepsilon$}
\AxiomC{$\Gamma, y:B, \Gamma' \vdash u:C$}
\RightLabel{\scriptsize{$ \oplus _e$}}
\TrinaryInfC{$\Gamma, \Delta, \Gamma' \vdash \delta (v, x.t, y.u)^\varepsilon:C$}
\DisplayProof
\vRuleSpace

\AxiomC{$x:A, \Gamma \vdash t:B$}
\RightLabel{\scriptsize{$\LmapSym_i$}}
\UnaryInfC{$\Gamma \vdash \lambda x.t: \Lmap{A}{B}$}
\DisplayProof
\hfil
\AxiomC{$\Gamma \vdash w: A$}
\AxiomC{$\Delta \vdash v: \Lmap{A}{B_\varepsilon}$}
\RightLabel{\scriptsize{$\LmapSym_e$}}
\BinaryInfC{$\Gamma, \Delta \vdash (v w)^\varepsilon: B$}
\DisplayProof
\vRuleSpace

\AxiomC{$\Gamma \vdash t:A$}
\AxiomC{$\Gamma \vdash u:B$}
\RightLabel{\scriptsize{$\With_i$}}
\BinaryInfC{$\Gamma \vdash \langle t, u \rangle: A\With B$}
\DisplayProof
\hfil
\AxiomC{$\Gamma \vdash v:A_\varepsilon\With B$}
\RightLabel{\scriptsize{$\With_{e1}$}}
\UnaryInfC{$\Gamma \vdash (\pi_1v)^\varepsilon:A$}
\DisplayProof
\AxiomC{$\Gamma \vdash v:A\With B_\varepsilon$}
\RightLabel{\scriptsize{$\With_{e2}$}}
\UnaryInfC{$\Gamma \vdash (\pi_2v)^\varepsilon:B$}
\DisplayProof

\end{framed}
\vspace*{-1.3em}
\caption{Typing rules for \LinLang}
\label{fig:typing}
\end{figure}

\mypar{Grammar of terms}

We define \emph{expressions} and \emph{values} with the following grammars:
\begin{align*}
\textrm{Expressions } t, u & \Coloneq v \mid (\Let x^+=t \In u)^+ \mid (\Let x^-=v \In u)^+ \mid \delta (v, (x,y).t)^+ \mid \\
 & \hphantom{{}\Coloneq{}} \delta (v, ().t)^+ \mid \delta (v, x.t, y.u)^+ \mid (v w)^+ \mid (\pi_1v)^+ \mid (\pi_2v)^+ \\
\textrm{Values } v, w & \Coloneq (\Let x^+=t \In v)^- \mid (\Let x^-=v \In w)^- \mid \delta (v, (x,y).w)^- \mid \\
 & \hphantom{{}\Coloneq{}} \delta (v, ().w)^- \mid \delta (v, x.w, y.w')^- \mid (v w)^- \mid (\pi_1v)^- \mid (\pi_2v)^- \mid \\
 & \hphantom{{}\Coloneq{}} x \mid \New \mid \Delete \mid (v, w) \mid () \mid \iota_1v \mid \iota_2v \mid \lambda x.t \mid \langle t, u \rangle \mid r_{n \in \mathbb{N}}
\end{align*}
Our untyped expressions do not have typing annotations. Instead,
expressions corresponding to elimination rules and let bindings have
polarity annotations, which is the minimal amount of information
determined by the type that we need to know the evaluation strategy.

Values are substitutable expressions: they are made of variables, all
negative expressions (as they follow the call-by-name evaluation
strategy) and positive expressions that have the shape of values in
call-by-value. Positive values are eliminated with pattern matching:
we note these with a dedicated $\delta$ eliminators for units, strict
pairs and sums. Negative value eliminators follow the usual notation
from lambda-calculus: lazy pair projections and function applications.

\mypar{Typing rules}
\label{typing-rules}

We define in \cref{fig:typing} typing rules for $\LinLang$.
Type polarity annotations in subscript $A_\varepsilon$ assert that $A$
has polarity $\varepsilon$. Contexts $\Gamma, \Delta$ are lists of
typed variables. $\Sigma(\Gamma, \Gamma')$ is the set of maps from
$\Gamma$ to $\Gamma'$ made of permutations and renamings. The term
$t[\sigma]$ appearing in the \emph{struct}(ural) rule denotes $t$
where all variables have been substituted according to $\sigma$; in
effect (struct) contains the unrestricted exchange rule. For
$\OLinLang$ we simply restrict $\sigma$ to be order-preserving.

Polarity annotations of expressions are inferred from the type
polarities in typing rules. Hence, we will leave them implicit for
typed expressions. The following notations allow to define and type
expressions without restrictions with respect to values, by picking an
arbitrary order to evaluate expressions. Such expressions
may not always be well-typed in \OLinLang, since let bindings can
only bind the right-most variable.
\begin{align*}
\iota_it^* &\eqdef \Let x=t \In \iota_ix & \delta^* (t, (x,y).u) &\eqdef \Let z=t \In \delta (z, (x,y).u) \\
(t, u)^* &\eqdef \Let x=t \In \Let y=u \In (x, y) & \delta^* (t, x.u, y.u') &\eqdef \Let z=t \In \delta (z, x.u, y.u') \\
(v t)^* &\eqdef \Let x=t \In v x & t; u &\eqdef \Let x=t \In \delta (x, ().u)
\end{align*}
As an example, we can define in \OLinLang the following program that
allocates then frees two resources. Upon allocation failure of $s$, it
is forced to free $r$ before returning:
\begin{align}
\vdash \delta^*\big(\New \Paren{}, r. \delta^*\Paren{\New \Paren{}, s.\Paren{\Delete s; \Delete r}, i.\Paren{i;\Delete r}}, i.i\big): 1
\label{example-two-resources}
\end{align}

\mypar{Purely-positive types without resources}
\label{sec:central-types}

An important notion that we will need is that of \emph{purely-positive
types without resources} defined as follows:
\begin{equation}
W \Coloneq 1 \mid W \otimes W' \mid W \oplus W' \label{def:central-type}
\end{equation}
These types are in particular \emph{central}: they commute for
$\otimes$ with any other types, naturally so. We indeed can define by
induction values \[\Swap^A_W: \Lmap{A \otimes W}{W \otimes A}\]
(corresponding to natural isomorphisms in the semantics
of \cref{sec:cat-model})\ifbool{arxiv}{:
\begin{align*}
\ifbool{arxiv}{
\Swap^A_I &\eqdef \lambda p. \delta\Paren{p, (a, i). \delta\Paren{i, ().\Paren{\star, a}}} \displaybreak[2]\\
}{}
\Swap^A_{W_1 \otimes W_2} &\eqdef \lambda p. \delta\Paren{p, (a, w). \delta\Paren{w, \Paren{x, x'}.t_{x,x'}}} \text{ where}\\
& \mspace{-70mu}t_{x,x'}=\Paren{\Let p_1 = \Swap^{A \otimes W_1}_{W_2} ((a, x), x') \In \delta\Paren{p_1, (y', p). \delta\Paren{p, (a, y). u_{y,y'}}}} \\
& \mspace{-70mu}u_{y,y'}=\Paren{\Let p_2 = \Swap^{W_2 \otimes A}_{W_1} ((y', a), y) \In \delta\Paren{p_2, (z, p). \delta\Paren{p, (z', a). \Paren{(z, z'), a}}}} \displaybreak[2]
\ifbool{arxiv}{
\\
\Swap^A_{W_1 \oplus W_2} &\eqdef \lambda p. \delta\Paren{p, (a, w). \delta\Paren{w,x.t_x,y.u_y}} \text{ where} \\
& \mspace{-70mu}t_x=\Paren{\Let p_1 = \Swap^A_{W_1} (a, x) \In \delta\Paren{p_1, (x', a). \Paren{\iota_1 x', a}}} \\
& \mspace{-70mu}u_y=\Paren{\Let p_2 = \Swap^A_{W_2} (a, y) \In \delta\Paren{p_2, (y', a). \Paren{\iota_2 y', a}}}
}{}
\end{align*}}{. }
These types also happen to be discardable ($\Lmap{A}{1}$) and copyable
($\Lmap{A}{A\otimes A}$) but we only rely on centrality in what follows.

\begin{figure}[tb]
\begin{framed}
\vspace*{0.5ex}
\hspace*{-0.7em}{\def\mymid{\mkern1mu{\mid}\mkern1mu}
$\begin{aligned}
\langle (\Let x^-=v \In t)^\varepsilon\mymid s\mymid l \rangle^\varepsilon &\rightsquigarrow \langle t[v/x]\mymid s\mymid l \rangle^\varepsilon \\
\langle (\Let x^+=t \In u)^\varepsilon\mymid s\mymid l \rangle^\varepsilon &\rightsquigarrow \langle t\mymid (x^+.u)^\varepsilon \cdotp s\mymid l \rangle^+ & \langle v\mymid (x^+.t)^\varepsilon \cdotp s\mymid l \rangle^+ &\rightsquigarrow \langle t[v/x]\mymid s\mymid l \rangle^\varepsilon \\
\langle (v w)^\varepsilon\mymid s\mymid l \rangle^\varepsilon &\rightsquigarrow \langle v\mymid w^\varepsilon\cdotp s\mymid l \rangle^- & \langle \lambda x.t\mymid v^\varepsilon \cdotp s\mymid l \rangle^- &\rightsquigarrow \langle t[v/x]\mymid s\mymid l \rangle^\varepsilon \\
\langle (\pi_i v)^\varepsilon\mymid s\mymid l \rangle^\varepsilon &\rightsquigarrow \langle v\mymid \pi_i^\varepsilon\cdotp s\mymid l \rangle^- & \langle \langle t_1, t_2 \rangle\mymid \pi_i^\varepsilon \cdotp s\mymid l \rangle^- &\rightsquigarrow \langle t_i\mymid s\mymid l \rangle^\varepsilon \\[1.5ex]
\langle \delta ((v,w), (x,y).t)^\varepsilon\mymid s\mymid l \rangle^\varepsilon &\rightsquigarrow \langle t[v/x, w/y]\mymid s\mymid l \rangle^\varepsilon & \langle \New\mymid ()\cdotp s\mymid \Nil \rangle^- &\rightsquigarrow \langle \iota_2()\mymid s\mymid \Nil \rangle^+ \\
\langle \delta ((), ().t)^\varepsilon\mymid s\mymid l \rangle^\varepsilon &\rightsquigarrow \langle t\mymid s\mymid l \rangle^\varepsilon & \mspace{-6mu}\langle \New\mymid ()\cdotp s\mymid r_n\cons l \rangle^- &\rightsquigarrow \langle \iota_1r_n\mymid s\mymid l \rangle^+ \\
\langle \delta (\iota_iv, x_1.t_1, x_2.t_2)^\varepsilon\mymid s\mymid l \rangle^\varepsilon &\rightsquigarrow \langle t_i[v/x_i]\mymid s\mymid l \rangle^\varepsilon & \langle \Delete\mymid r_n\cdotp s\mymid l \rangle^- &\rightsquigarrow \langle ()\mymid s\mymid r_n\cons l \rangle^+
\end{aligned}$\vspace*{0.5ex}}\hspace*{-0.7em}\end{framed}
\vspace*{-1.3em}
\caption{Reduction rules}
\label{fig:reductions}
\end{figure}

\mypar{Operational semantics}

We define in \cref{fig:reductions} a small-step operational semantics for
untyped expressions with an abstract machine. We begin by defining a
stack that stores the arguments of delayed operations along with their
polarity: indices of lazy pair projections, arguments of function
applications and continuations of let bindings of positive values.
Finally, a command consists of an expression, a stack, a free-list
of resources interpreting the allocation monad and the current polarity.
\begin{align*}
\textrm{Stacks } s &\Coloneq \star \mid v^\varepsilon\cdotp s \mid \pi_i^\varepsilon\cdotp s \mid (x^+.u)^\varepsilon\cdotp s & \\
\textrm{Lists } l &\Coloneq \Nil \mid r_{n \in \mathbb{N}}\cons l & \\
\textrm{Commands } c &\Coloneq \langle t\mid s\mid l \rangle^\varepsilon &
\end{align*}
We now define reduction rules: let bindings of negative expressions
perform the substitution them immediately, following call-by-name
reduction, while let bindings of positive expressions push their
continuation on the stack to first reduce the expression to a value.
All rules that push arguments on the stack come with their dual rule
that consumes the argument on the stack. Pattern-matching rules can
reduce immediately as only values can be bound. Finally, rules for
constants $\New$ and $\Delete$ are the only ones that manipulate the
free-list of resources.

If we interpret our previous program $p=(\ref{example-two-resources})$
with the list of resources $r_0\cons r_1\cons l$, we obtain in
particular the following reduction steps:
\begin{equation*}
\begin{split}
    &\langle p \mid \star \mid r_0\cons r_1\cons l \rangle^+ \\
    &\rightsquigarrow^* \Bigmachine{2}{\iota_1r_1}{\Paren{x^+.\delta \Paren{x, s.\Paren{\Delete s; \Delete r_0}, i.\Paren{i;\Delete r_0}}} \cdotp \star}{l}^+ \\
    &\rightsquigarrow^* \langle \Paren{} \mid \star \mid r_0\cons r_1\cons l \rangle^+
\end{split}
\end{equation*}
\vspace{-2ex}

\begin{figure}[tb]
\begin{framed}
\centering

\AxiomC{}
\RightLabel{}
\UnaryInfC{$\star: A \vdashp A$}
\DisplayProof
\hfil
\AxiomC{$s: B_\varepsilon \vdashp C$}
\AxiomC{$\vdashp v:A$}
\RightLabel{}
\BinaryInfC{$v^\varepsilon \cdotp s: \Lmap{A}{B} \vdashp C$}
\DisplayProof
\hfil
\AxiomC{$s: B_\varepsilon \vdashp C$}
\AxiomC{$x: A \vdashp t: B$}
\RightLabel{}
\BinaryInfC{$(x^+.t)^\varepsilon \cdotp s: A \vdashp C$}
\DisplayProof
\vfil
\vRuleSpace
\AxiomC{$s: A_\varepsilon \vdashp C$}
\RightLabel{}
\UnaryInfC{$\pi_1^\varepsilon \cdotp s: A \With B \vdashp C$}
\DisplayProof
\hfil
\AxiomC{$s: B_\varepsilon \vdashp C$}
\RightLabel{}
\UnaryInfC{$\pi_2^\varepsilon \cdotp s: A \With B \vdashp C$}
\DisplayProof

\end{framed}
\vspace*{-1.3em}
\caption{Stack typing rules}
\label{fig:stack}
\end{figure}

\mypar{Properties}

We now study properties of \LinLang. First, we establish standard
properties such as confluence (through determinism), subject reduction
and progress.

\begin{proposition}\label{prop:determinism}
The reduction rules are deterministic: at most one reduction rule can
be applied to any command\ifbool{arxiv}{ (see \cref{proof:determinism})}{}.
\end{proposition}

We define typing judgements for expressions $\vdashp t:
A$ by extending all typing rules given in \cref{typing-rules} with the
following axiom scheme for resources: $\vdashp r_n: R$, to type resources
that occur in commands through a program execution. We then extend
those judgements in \cref{fig:stack} for closed stacks $s: A \vdashp B$ which follow
sequent calculus rules. Finally, we extend judgements to commands
$c:A$, which consist of pairs of judgements $\vdashp t: B_\varepsilon$
and $s: B \vdashp A$ for $c = \langle t \mid s \mid
l \rangle^\varepsilon: A$.
We prove the following substitution lemma by induction on the derivation of $t$:

\begin{lemma}[Substitution lemma (SL)]
If $\vdashp v:A$ and $\Gamma, x:A, \Gamma' \vdashp t:B$, then
$\Gamma, \Gamma' \vdashp t[v/x]:B$.
\end{lemma}

\begin{theorem}[Subject reduction]\label{thm:subject-reduction}
The reduction rules preserve typing judgements: for any $c_1:A$ and
$c_2$, $c_1 \rightsquigarrow c_2$ implies $c_2: A$. This also holds in
the ordered fragment, i.e.\ for $\vdashp$ without the exchange
rule.
\ifbool{arxiv}{(See \cref{proof:subject-reduction}.)}{}
\end{theorem}

We define final values $v_t \Coloneq () \mid (v,
w) \mid \iota_iv \mid r_n \mid \langle t, u \rangle \mid \lambda
x.t \mid \New \mid \Delete$. They include in particular all values
$v$ such that $\vdashp v: A_+$.

\begin{theorem}[Progress]\label{thm:progress}
A well-typed command reduces if and only if it is not of the shape
$\langle v_t\mid \star\mid l \rangle^\varepsilon$. This also holds
in \OLinLang. \ifbool{arxiv}{(See \cref{proof:progress}.)}{}
\end{theorem}

Together, subject reduction and progress ensures that any well-typed
expression $\vdashp t:A$ reduces either indefinitely\footnote{Although
we do not have non-terminating features, our proof by progress \&
subject reduction does not rely on termination.} or to a final command
$\langle v_t\mid \star\mid l \rangle$ with $\vdashp v_t:A$.

\section{Resource-safety properties}
\label{sec:resource-props}

In this section, we prove a resource-safety property for \LinLang
pertaining to its linear character. This property is stated for
\emph{complete programs}: closed expressions that return a purely positive
type without resource, executed in an empty context. We show that for
such a program $t$, the execution starting with any free-list leaves
it unchanged upon return, up to a permutation $\sigma$:
\begin{align*}
\langle t\mid \star\mid
l \rangle &\rightsquigarrow^* \langle v\mid \star\mid l' \rangle
&&\Longrightarrow&\exists \sigma, l'&=\sigma(l)
\end{align*}
The property is indeed about resource safety as it provides several
good properties expected in programs that manipulate resources:
\begin{itemize}
\item all allocated resources are released by the end of the program,
\item only previously-allocated resources are released,
\item no resource has been released twice (i.e. no ``use after free'',
in this limited context where the only way to use a resource is to
release it),
\item these properties remain true in case of an error during the
program execution, including an allocation error.
\end{itemize}
We start by describing a stronger property for \OLinLang: in the
ordered language, the final free-list is identical to the initial one
($\sigma=\operatorname{id}$). This corresponds to the expected
property with ordered type systems that resources are freed in a LIFO
order~\cite[§1.4]{Walker2005a}. We also provide a counter-example
showing that we cannot include right functions in \OLinLang without
breaking the LIFO property.

\mypar{Ordered case}

Let $W$ a purely positive type without resource and $\vdash t: W$ a
closed typed expression of $\OLinLang$ (i.e. without exchange rule).

\begin{proposition}\label{prop:ordp}
For any value $v$ and lists of resources $l,l'$,
if \[\langle t\mid \star\mid l \rangle \rightsquigarrow^* \langle
v\mid \star\mid l' \rangle\] then $l'=l$.
\end{proposition}

To prove this property, we will define the list of resources $LR(t)$
of a term $t$ by tracking resources in contexts of typing rules, and
show that such lists are invariant by reduction. In order to reason
about resources in case of substitutions, $LR(t[v/x])$ must be
definable in terms of $LR(t)$ and $LR(v)$, hence their concatenation.
To ensure this (see \cref{lem:ord-sl}), we restrict contexts to have the shape
$\Theta \eqdef \Gamma; L; \Delta$ with $L$ the list of resources,
$\Gamma, \Delta$ contexts of variables and $x$ to be the right-most
variable in $\Gamma$ or the left-most variable in $\Delta$.

Given $\Concat$ the concatenation operation for lists,
$\Theta = \Gamma_\Theta;L_\Theta;\Delta_\Theta$ and $\Theta'
= \Gamma_{\Theta'};L_{\Theta'};\Delta_{\Theta'}$, the expression
$\Theta\CtxComp\Theta'$ asserts that we are in one of the three
following cases to define the concatenation of both contexts:
\begin{itemize}
    \item $L_\Theta = []$, then
    $\Theta\CtxComp\Theta' \eqdef \Gamma_\Theta, \Delta_\Theta, \Gamma_{\Theta'};L_{\Theta'};\Delta_{\Theta'}$.
\item $L_{\Theta'} = []$, then
    $\Theta\CtxComp\Theta' \eqdef \Gamma_\Theta;L_\Theta;\Delta_\Theta, \Gamma_{\Theta'}, \Delta_{\Theta'}$.
\item $\Delta_\Theta = \Gamma_{\Theta'} = \emptyset$, then
    $\Theta\CtxComp\Theta' \eqdef \Gamma_\Theta; L_\Theta \Concat
    L_{\Theta'}; \Delta_{\Theta'}$.
\end{itemize}

\begin{figure}[tb]
\begin{framed}
\centering
    
\AxiomC{}
\RightLabel{}
\UnaryInfC{$;[];x\vdasho x$}
\DisplayProof
\hfil
\AxiomC{}
\RightLabel{}
\UnaryInfC{$;[];\vdasho ()$}
\DisplayProof
\hfil
\AxiomC{}
\RightLabel{}
\UnaryInfC{$;[];\vdasho \New$}
\DisplayProof
\hfil
\AxiomC{}
\RightLabel{}
\UnaryInfC{$;[];\vdasho \Delete$}
\DisplayProof
\hfil
\AxiomC{}
\RightLabel{}
\UnaryInfC{$;[r_n]; \vdasho r_n$}
\DisplayProof

\vfil
\vRuleSpace

\AxiomC{$\Gamma;[];x, \Delta \vdasho x$}
\RightLabel{}
\UnaryInfC{$\Gamma, x;[];\Delta \vdasho x$}
\DisplayProof
\hfil
\AxiomC{$\Gamma, x;[];\Delta \vdasho x$}
\RightLabel{}
\UnaryInfC{$\Gamma;[];x, \Delta \vdasho x$}
\DisplayProof

\vfil
\vRuleSpace

\AxiomC{$\Theta\vdasho v$}
\RightLabel{}
\UnaryInfC{$\Theta\vdasho \iota_iv$}
\DisplayProof
\hfil
\AxiomC{$\Theta\vdasho v$}
\RightLabel{}
\UnaryInfC{$\Theta\vdasho \pi_iv$}
\DisplayProof
\hfil
\AxiomC{$\Theta\vdasho t$}
\AxiomC{$\Theta\vdasho u$}
\RightLabel{}
\BinaryInfC{$\Theta\vdasho \langle t, u \rangle$}
\DisplayProof
\hfil
\AxiomC{$x,\Theta\vdasho t$}
\RightLabel{}
\UnaryInfC{$\Theta\vdasho \lambda x.t$}
\DisplayProof

\vfil
\vRuleSpace

\AxiomC{$\Theta\vdasho v$}
\AxiomC{$\Theta'\vdasho w$}
\RightLabel{}
\BinaryInfC{$\Theta\CtxComp\Theta'\vdasho (v, w)$}
\DisplayProof
\hfil
\AxiomC{$\Theta,x\vdasho u$}
\AxiomC{$\Theta'\vdasho t$}
\RightLabel{}
\BinaryInfC{$\Theta\CtxComp\Theta'\vdasho \Let x = t \In u$}
\DisplayProof

\vfil
\vRuleSpace

\AxiomC{$\Theta,x,y \CtxComp \Theta''\vdasho t$}
\AxiomC{$\Theta'\vdasho v$}
\RightLabel{}
\BinaryInfC{$\Theta\CtxComp\Theta'\CtxComp\Theta''\vdasho \delta (v, (x,y).t)$}
\DisplayProof
\hfil
\AxiomC{$\Theta \CtxComp \Theta''\vdasho t$}
\AxiomC{$\Theta'\vdasho v$}
\RightLabel{}
\BinaryInfC{$\Theta\CtxComp\Theta'\CtxComp\Theta''\vdasho \delta (v, ().t)$}
\DisplayProof

\vfil
\vRuleSpace

\AxiomC{$\Theta,x \CtxComp \Theta''\vdasho t$}
\AxiomC{$\Theta,y \CtxComp \Theta''\vdasho u$}
\AxiomC{$\Theta'\vdasho v$}
\RightLabel{}
\TrinaryInfC{$\Theta\CtxComp\Theta'\CtxComp\Theta''\vdasho \delta (v, x.t, y.u)$}
\DisplayProof
\hfil
\AxiomC{$\Theta\vdasho w$}
\AxiomC{$\Theta'\vdasho v$}
\RightLabel{}
\BinaryInfC{$\Theta\CtxComp\Theta'\vdasho v w$}
\DisplayProof

\end{framed}
\vspace*{-1.3em}
\caption{Typing rules of ordered expressions with resources (types omitted)}
\label{fig:ordered-rules}
\end{figure}

\begin{figure}[tb]
\begin{framed}
\centering

\AxiomC{}
\RightLabel{}
\UnaryInfC{$[] \vdasho^S \star$}
\DisplayProof
\hfil
\AxiomC{$;L_v; \vdasho v$}
\AxiomC{$L_s \vdasho^S s$}
\RightLabel{}
\BinaryInfC{$L_s \Concat L_v \vdasho^S v\cdotp s$}
\DisplayProof

\vfil
\vRuleSpace

\AxiomC{$;L_t; x \vdasho t$}
\AxiomC{$L_s \vdasho^S s$}
\RightLabel{}
\BinaryInfC{$L_s \Concat L_t \vdasho^S (x^+.t)\cdotp s$}
\DisplayProof
\hfil
\AxiomC{$;L_t; \vdasho t$}
\AxiomC{$L_s \vdasho^S s$}
\RightLabel{}
\BinaryInfC{$L_s \Concat L_t \Concat l \vdasho^C \langle t \mid s \mid l \rangle$}
\DisplayProof

\end{framed}
\vspace*{-1.3em}
\caption{Typing rules of ordered stacks and commands with resources (types omitted)}
\end{figure}

We can now type terms with resources with the judgement $\vdasho$
indexed by such contexts, which enrich the previous typing judgements
with information to track their resources
(in \cref{fig:ordered-rules}, where types are omitted for brevity).
Given an ordered expression $\Gamma; L; \Delta \vdasho t$, we define
its list of resources $LR(t) \eqdef L$. For $\Theta = \Gamma;
L; \Delta$, we define the notations $\Theta, x \eqdef \Gamma;
L; \Delta, x$ and $x, \Theta \eqdef x, \Gamma; L; \Delta$.

\begin{lemma}\label{lem:vdasho}
If $\Gamma \vdash t:A$, then $\Gamma;; \vdasho t:A$\ifbool{arxiv}{ (see~\cref{proof:vdasho})}{}.
\end{lemma}

We can then accept substitutions of expressions that preserve
well-formed contexts. The following left and right substitution lemmas
cover substitutions encountered in the operational semantics.

\begin{lemma}
\label{lem:ord-sl}
\emph{(Left SL)} If $;L_t; \vdasho t$ and $\Gamma, x; L_u; \Delta \vdasho u$,
then $\Gamma; L_t \Concat L_u; \Delta \vdasho u[t/x]$.
\emph{(Right SL)} If $;L_t; \vdasho t$ and $\Gamma; L_u;
x, \Delta \vdasho u$, then $\Gamma; L_u \Concat L_t; \Delta \vdasho
u[t/x]$\ifbool{arxiv}{ (see \cref{proof:ord-sl})}{}.
\end{lemma}

We then extend $\vdasho$ for stacks and commands, with only resources
in their contexts. Resources from stacks remain to the left of resources from
expressions in the context of commands:

\begin{theorem}\label{thm:ord-inv}
Reducing an ordered command results in an ordered command with the
same list of resources, i.e. for all $c_1,c_2,M$ such that
$L \vdasho^C c_1$ and $c_1 \rightsquigarrow c_2$ one has $L \vdasho^C
c_2$\ifbool{arxiv}{ (see \cref{proof:ord-inv})}{}.
\end{theorem}

We can now prove \cref{prop:ordp}:

\begin{proof}[\cref{prop:ordp}]
For any expression with a purely-positive type without resource
$\vdash t: W$ in \OLinLang, by subject reduction and progress we have
that $\langle t\mid \star\mid l \rangle$ either reduces indefinitely,
or there exists a final value $\vdash v_t: W$ such that $\langle
t\mid \star\mid l \rangle \rightsquigarrow^* \langle v\mid \star\mid
l' \rangle$.
We then prove by induction on derivations that typed expressions
without structural rules are ordered: they do not include resources,
so concatenated contexts are always well-formed. So both $t$ and $v$
are ordered and without resources. Since reduction rules preserve the
list of resources, we have $l=l'$.
\end{proof}

\mypar{A counter-example for right functions}

We only have left functions in \OLinLang, that is, abstractions
binding the leftmost variable together with the matching elimination
rule from ordered logic. We provide the following counter-example to
the resource-safety property with right functions. We underline the
right function abstractions and applications.
\begin{align}
p & \eqdef \delta^*\Paren{\New (), r. \delta^*\Paren{\New (), s.\Paren{\underline{t_r s}}, i.\Paren{\Delete r; i}}, i.i} \nonumber\\
&
\mathrel{\hphantom{\eqdef}}\text{where }t_r=\Paren{\Let i = \Delete r \In \delta\Paren{i, ().\underline{\lambda x}.\Delete x}}^-\label{def-incorrect-term}
\end{align}
This example does not belong to \OLinLang as it uses the right
function $t_r:R\multimap 1$ and its application $\underline{t_rs}$.
\begin{center}
{
\def\fCenter{\vdash}
\AxiomC{}
\doubleLine
\RightLabel{\scriptsize{$\LmapSym_e$}}
\UnaryInf$r:R \fCenter \Delete r : 1$
\AxiomC{}
\doubleLine
\RightLabel{\scriptsize{$\LmapSym_e$}}
\UnaryInf$x:R \fCenter \Delete x : 1$
\LeftLabel{\scriptsize{$\boldsymbol\times$}}
\RightLabel{\scriptsize{$\RmapSym_i$ (not in $\OLinLang$)}}
\UnaryInf$\fCenter \underline{\lambda x}.\Delete x : \Rmap{R}{1}$
\RightLabel{\scriptsize{$1_e$}}
\UnaryInf$i:1 \fCenter \delta\Paren{i, ().\underline{\lambda x}.\Delete x} : \Rmap{R}{1}$
\RightLabel{\scriptsize{$\Let$}}
\BinaryInf$r:R \fCenter \Let i = \Delete r \In \delta\Paren{i, ().\underline{\lambda x}.\Delete x} : \Rmap{R}{1}$
\RightLabel{\scriptsize{$=$}}
\dottedLine{}
\UnaryInf$\vphantom{X^{X^X}_{X_X}}r:R \fCenter t_r : \Rmap{R}{1}$
\noLine{}
\let\oldVskip\extraVskip
\def\extraVskip{-1.3ex}
\def\fCenter{\mathrel{\phantom{\vdash}}}
\UnaryInf$\phantom{r:R }\fCenter\phantom{t_r : \Rmap{R}{1}}$
\let\extraVskip\oldVskip
\def\fCenter{\vdash}
\AxiomC{}
\UnaryInf$s:R \fCenter s:R$
\LeftLabel{\scriptsize{$\boldsymbol\times$}}
\RightLabel{\scriptsize{$\RmapSym_e$ (not in $\OLinLang$)}}
\insertBetweenHyps{\hspace{-12em}}
\BinaryInf$r:R, s:R \fCenter \underline{t_r s} : 1$
\DisplayProof
}
\end{center}
As we can see from the typing derivation, the context is ordered as
$r,s$, which is the converse of what we would have obtained with a
left function. Subsequently, we can then calculate that two resources
$r_0$ and $r_1$ are reordered during the execution as follows (with
the obvious reductions rules for the right abstraction and
application):
\[
\langle p  \mid \star \mid r_0\cons r_1\cons l \rangle \rightsquigarrow^* \langle () \mid \star \mid r_1\cons r_0\cons l \rangle
\]
This contradicts the LIFO property for \OLinLang.

\mypar{Non-ordered case}
If we consider expressions $\vdash t: W$ in \LinLang, that is by
adding the exchange rule, they have a weaker property:

\begin{theorem}\label{thm:linp}
For all $v,l,l'$ such that $\langle t\mid \star\mid
l \rangle \rightsquigarrow^* \langle v\mid \star\mid l' \rangle$,
there exists a permutation $\sigma$ of lists of resources such that
$l'=\sigma(l)$.
\end{theorem}

This is proved similarly to \cref{prop:ordp}, by
forgetting the order of variables and resources in the predicate
$\Gamma; L; \Delta \vdasho t$. This simplifies context concatenation
by removing preconditions and requires a single substitution lemma.
\ifbool{arxiv}{The proof is detailed in \cref{proof:linearity}.}{}

\section{The resource call-by-push-value}
\label{sec:affine-lang}

We now introduce a resource CBPV: an ordered variant of CBPV with an
allocation effect $\New$, exceptions, a chosen effectful destructor
$\DropB$ at all types, and a $\Move$ operation that implements a
side-effecting exchange rule.

Formally, we define a calculus \AffLang which extends \OLinLang. Its
semantics is given by translation into \LinLang : exceptions are
propagated as errors, removing variables from the context with their
associated destructor. Hence, \LinLang serves as a meta-language in
which the ambient allocation monad remains implicit, and which gives
resource-safety properties for \AffLang. In addition, by removing
$\Move$, we obtain a calculus $\OAffLang$ whose translation falls into
$\OLinLang$.

As a starting point, we picture the situation described
in \cref{sec:monoidal-destructor-cat}:
\begin{equation}\label{eq:adjunction-RCBPV}
\begin{tikzcd}[column sep = 2em]
\phantom{\Ccat}\mathllap{\Ccat_{/TI}} \arrow[rr,"U^\Dm", bend left] & \bot \mathrlap{\scriptstyle{\;(a)}} &
\Ccat\arrow[ll,"F^\Dm",bend left]\arrow[rr,"F^{T\Em}", bend left] & \bot \mathrlap{\scriptstyle{\;(b)}} &
\mathrlap{\Ccat^{T\Em}}\phantom{\Ccat}\arrow[ll,"U^{T\Em}",bend left]
\end{tikzcd}
\end{equation}
where $\Dm={-\With TI}$ provides, for an arbitrary linear type, the
free type with destructor (given by its second projection).

When trying to add exceptions to a general notion of effect given by a
strong monad $T$, one would look at the adjunction $(b)$, which,
assuming $T\Em$ strong, gives rise to a linear CBPV
model~\cite{CFM15}, with positive types interpreted in $\Ccat$ and
negative types interpreted in $\Ccat^{T\Em}$. It is unnatural, though,
to assume that $T\Em$ is strong. In order to recover a strength for
$T\Em$, we want to restrict the positive types to the linear types
that are provided with a chosen destructor, by looking at the monoidal
adjunction $(a)$ above. When starting from an adjunction model,
composing with a monoidal adjunction on the left yields another
adjunction model, a construction which generalises the Girard
translations~\cite[§5.3]{CFM15} (see
also~\cite{Munch-Maccagnoni2017ll-cbpv} for more details). This
suggests that we look at the adjoint situation
\begin{equation*}
\begin{tikzcd}[column sep = 2em]
\phantom{\Ccat}\mathllap{\Ccat_{/TI}} \arrow[rr,"\up", bend left] & \bot &
\mathrlap{\Ccat^{T\Em}}\phantom{\Ccat}\arrow[ll,"\down",bend left]
\end{tikzcd}
\end{equation*}
obtained by composition with $\up\eqdef F^{T\Em}U^\Dm$ and
$\down\eqdef F^\Dm U^{T\Em}$. Note that in terms of underlying types,
one has $\up A = T(A\oplus E)$ and $\down A=A\With TI$.

We cannot apply the results of~\cite{CFM15} directly, which only deals
with the situation where the monoidal categories are symmetric and the
monad is strong. Instead, we do this construction by hand, which we
give as a translation of the resource CBPV into
$\OLinLang$ and $\LinLang$. Within this more focused approach, we
refine the translation underlying \cite[§5.3]{CFM15} with ordered
sequents, and with the consideration of a strength of $T\Em$ with
respect to $U^\Dm$, which is necessary for having antecedents with
several variables in the first place.

Another important difference is that we are translating direct-style
calculi into direct-style calculi: the monad $T$ is, already, the
ambient monad for side-effects in $\OLinLang$ and $\LinLang$. As it
turns out, we can still adapt the translation to work in this way.
Note that the type $\down A=A\With TI$ is $A\With 1$ in $\OLinLang$,
and that the type $\up A = T(A\oplus E)$ describes effectful
computations of type $A\oplus E$ in $\OLinLang$. In what follows, we
make use of these type constructions in $\OLinLang$ which we note
$\Down A \eqdef A\With 1$ and $\Up A\eqdef A\oplus E$.

\mypar{Expressions and types}

\begin{figure}[tb]
\begin{framed}
\centering

\AxiomC{}
\RightLabel{\scriptsize{var}}
\UnaryInfC{$x:A \vdash x: A$}
\DisplayProof
\hfil
\AxiomC{$\Delta \vdash t: A$}
\AxiomC{$\Gamma, x:A \vdash u: B$}
\RightLabel{\scriptsize{$\Let$}}
\BinaryInfC{$\Gamma, \Delta \vdash \Let x=t \In u: B$}
\DisplayProof
\vRuleSpace

\AxiomC{}
\RightLabel{\scriptsize{$\DropB_A$}}
\UnaryInfC{$\vdash \DropB_A: \Lmap{A}{1}$}
\DisplayProof
\hfil
\AxiomC{$\Gamma, \Gamma', x:A \vdash t: C$}
\RightLabel{\scriptsize{$\Move$}}
\UnaryInfC{$\Gamma, x:A, \Gamma' \vdash \Move(x) \In t: C$}
\DisplayProof
\vRuleSpace

\AxiomC{}
\RightLabel{\scriptsize{$\New$}}
\UnaryInfC{$\vdash \New: \Lmap{1}{R}$}
\DisplayProof
\hfil
\AxiomC{}
\RightLabel{\scriptsize{$\RaiseB$}}
\UnaryInfC{$\vdash \RaiseB: \Lmap{E}{A}$}
\DisplayProof
\vRuleSpace

\AxiomC{$\Delta \vdash t: P$}
\AxiomC{$\Gamma, x:P \vdash u: A$}
\AxiomC{$\Gamma, e:E \vdash u': A$}
\RightLabel{\scriptsize{$\Try$}}
\TrinaryInfC{$\Gamma, \Delta \vdash \Try x \Leftarrow t \In u \Unless e \Rightarrow u': A$}
\DisplayProof
\vRuleSpace

\AxiomC{$\Gamma \vdash v: A$}
\AxiomC{$\Delta \vdash w: B$}
\RightLabel{\scriptsize{$\otimes_i$}}
\BinaryInfC{$\Gamma, \Delta \vdash (v, w): A \otimes B$}
\DisplayProof
\hfil
\AxiomC{$\Delta \vdash v: A \otimes B$}
\AxiomC{$\Gamma, x:A, y:B, \Gamma' \vdash t: C$}
\RightLabel{{\scriptsize{$\otimes_e$}}\hspace*{-1em}}
\BinaryInfC{$\Gamma, \Delta, \Gamma' \vdash \delta (v, (x,y).t): C$}
\DisplayProof
\vRuleSpace

\AxiomC{}
\RightLabel{\scriptsize{$1_i$}}
\UnaryInfC{$\vdash (): 1$}
\DisplayProof
\hfil
\AxiomC{$\Delta \vdash v: 1$}
\AxiomC{$\Gamma, \Gamma' \vdash t: A$}
\RightLabel{\scriptsize{$1_e$}}
\BinaryInfC{$\Gamma, \Delta, \Gamma' \vdash \delta (v, ().t): A$}
\DisplayProof
\vRuleSpace

\AxiomC{$\Gamma \vdash v: A$}
\RightLabel{\scriptsize{$\oplus_{i1}$}}
\UnaryInfC{$\Gamma \vdash \iota_1v: A \oplus B$}
\DisplayProof
\hfil
\AxiomC{$\Gamma \vdash v:B$}
\RightLabel{\scriptsize{$\oplus_{i2}$}}
\UnaryInfC{$\Gamma \vdash \iota_2v: A \oplus B$}
\DisplayProof
\vRuleSpace

\AxiomC{$\Delta \vdash v: A \oplus B$}
\AxiomC{$\Gamma, x:A, \Gamma' \vdash t: C$}
\AxiomC{$\Gamma, y:B, \Gamma' \vdash u: C$}
\RightLabel{\scriptsize{$\oplus_e$}}
\TrinaryInfC{$\Gamma, \Delta, \Gamma' \vdash \delta(v, x.t, y.u): C$}
\DisplayProof
\vRuleSpace

\AxiomC{$x:A, \Gamma \vdash t: B$}
\RightLabel{\scriptsize{$\LmapSym_i$}}
\UnaryInfC{$\Gamma \vdash \lambda x.t: \Lmap{A}{B}$}
\DisplayProof
\hfil
\AxiomC{$\Gamma \vdash w: A$}
\AxiomC{$\Delta \vdash v: \Lmap{A}{B}$}
\RightLabel{\scriptsize{$\LmapSym_e$}}
\BinaryInfC{$\Gamma, \Delta \vdash v w: B$}
\DisplayProof
\vRuleSpace

\AxiomC{$\Gamma \vdash t: A$}
\AxiomC{$\Gamma \vdash u: B$}
\RightLabel{\scriptsize{$\With_i$}}
\BinaryInfC{$\Gamma \vdash \langle t, u \rangle: A \With B$}
\DisplayProof
\hfil
\AxiomC{$\Gamma \vdash v: A \With B$}
\RightLabel{\scriptsize{$\With_{e1}$}}
\UnaryInfC{$\Gamma \vdash \pi_1 v: A$}
\DisplayProof
\hfil
\AxiomC{$\Gamma \vdash v: A \With B$}
\RightLabel{\scriptsize{$\With_{e2}$}}
\UnaryInfC{$\Gamma \vdash \pi_2 v: B$}
\DisplayProof
\vRuleSpace

New and changed rules are in \textbf{bold}.

\end{framed}
\vspace*{-1.3em}
\caption{Typing rules for \AffLang}
\end{figure}

The grammar of terms of \AffLang extends that for \OLinLang, with the
following terms:
\begin{itemize}
  \item ``$\New$'', now of type $\Lmap R 1$, which raises an exception in case of allocation error.
  \item ``$\DropB_A$'', a function of type $\Lmap A 1$ that releases the resources contained in its argument.
  \item ``$\Move(x) \In t$'', an expression that places $x$ on top of the stack in $t$.
  \item ``$\RaiseB$'', a value that inhabits any type with a given exception.
  \item ``$\Try x \Leftarrow t \In u \Unless e \Rightarrow u'$'', an expression that catches exceptions occuring in $t$ based on \cite{Benton2001}.
\end{itemize}
We call \OAffLang the fragment of \AffLang without $\Move$.
Also, from now on we leave polarity annotations implicit because they
can always be inferred from the type.

The grammar of types is unchanged:
\[
\text{Types }A,B:\begin{cases}
\text{Positive types } & P,Q \Coloneq R \mid 1 \mid A \otimes B \mid A \oplus B \\
\text{Negative types } & N,M \Coloneq \Lmap{A}{B} \mid A \With B \\
\end{cases}
\]
However, the interpretation of types is changed, since each positive
type is assigned a chosen destructor.

The calculus is parameterised by a type of exceptions $E$. In
$\OAffLang$, exceptions must not exchange resources during stack
unwinding, so we add the constraint that \emph{$E$ is a
purely-positive type without resource}. We also assume given some
closed value $\vdash \NewFail: E$, which is the exception raised
whenever an allocation fails. (For instance, $E=1$ and $\NewFail=()$.)
We will rely on terms $\Swap^A_W$ in \OAffLang defined as
in \OLinLang (\cref{sec:central-types}).

\mypar{Move as an effectful exchange rule}

As explained in \cref{sec:monoidal-destructor-cat}, the category $\Ccat_{/TI}$ in the situation (\ref{eq:adjunction-RCBPV})
depicted above is not symmetric in general. However, when $\Ccat$ is
symmetric, there is nevertheless an effectful map $A\otimes
B\rightarrow B\otimes A$, as found in the following hom-set:
\[
\Ccat_{/TI}(A\otimes B,F^\Dm U^\Dm(B\otimes A)) \cong
\Ccat(U^\Dm A\otimes U^\Dm B,U^\Dm B\otimes U^\Dm A)
\]
This is reflected in $\AffLang$ with the operation $\Move$ which
exchanges variables. It is not presented as a structural rule in the
traditional way: it indeed performs an effect and does not commute
with other rules.
In the following example, two terms allocate three resources and then
frees them in reverse order. They would be identified if $\Move$ was
treated as a structural rule in the usual manner (e.g. as in
$\LinLang$), but they behave differently:
\begin{align*}
&\begin{aligned}
&\Let r = \New () \In \\
&\Let s = \New () \In \\
&\Let t = \New () \In \\
&\DropB_R t; \DropB_R s; \DropB_R r
\end{aligned}
&&\begin{aligned}
&\Let r = \New () \In\\
&\Let s = \New () \In \Move(r) \In\\
&\Let t = \New () \In \\
&\DropB_R t; \Move(s) \In \DropB_R s; \DropB_R r
\end{aligned}
\end{align*}
Both terms allocate three resources then frees them in reverse order.
If the allocation of $t$ fails, then the raised exception will free
the first two resources in a different order, which can be observed
with the evaluation context $\mid\star\mid [r_0,r_1]\rangle$: the
final free-lists are respectively $[r_0,r_1]$ and $[r_1,r_0]$.

For this reason, $\Move$ cannot give rise to a \emph{contextual
isomorphism}~\cite{Levy2017} between $A\otimes B$ and $B \otimes A$
for arbitrary $A$ and $B$, that is, which would allow us to replace
$A\otimes B$ with $B \otimes A$ in arbitrary context without changing
the meaning.
Since a contextual isomorphism just requires inverses and purity
(thunkability)~\cite{Levy2017}, $\Move$ must be effectful in any
reasonable interpretation.

\mypar{Translation into \LinLang}

Each type $A$ has a positive
interpretation $A^+$ equipped with a destructor written
$\Drop_A$ and a negative interpretation $A^-$.
{
\allowdisplaybreaks
\begin{align*}
1^+ &\eqdef 1  &  \Drop_1 &\eqdef \lambda v.v \\
R^+ &\eqdef R  &  \Drop_R &\eqdef \lambda r.\Delete r \\
(A \otimes B)^+ &\eqdef A^+ \otimes B^+  &  \Drop_{(A \otimes B)} &\eqdef \lambda p.\delta\Paren{p, (a,b).\Paren{\Drop_B b; \Drop_A a}} \\
(A \oplus B)^+ &\eqdef A^+ \oplus B^+  &  \Drop_{(A \oplus B)} &\eqdef \lambda s.\delta\Paren{s, a.\Paren{\Drop_A a}, b.\Paren{\Drop_B b}} \\
N^+ &\eqdef \Down N^-  &  \Drop_N &\eqdef \lambda a. \pi_2 a \\
(\Lmap{A}{B})^- &\eqdef \Lmap{A^+}{B^-} & & \\
(A \With B)^- &\eqdef A^- \With B^- & \star^+ &\eqdef \star \\
P^- &\eqdef \Up P^+ & (\Gamma, x:A)^+ &\eqdef \Gamma^+, x:A^+
\end{align*}}Note that for any purely positive type (hence also $E$), we have $W^+
= W$.
We define three expressions which we use in the translation:
\begin{itemize}
\item $\Gamma^+ \vdash \DropCtx_\Gamma: 1$ by induction on $\Gamma$, which drops each variable from right to left.
\begin{align*}
    \DropCtx_{\star} &\eqdef () \\
    \DropCtx_{\Gamma, x:A} &\eqdef \Drop_A x; \DropCtx_\Gamma
\intertext{\item $\Gamma^+, e:E \vdash \Unwind_\Gamma(e): E$ by induction on $\Gamma$, which drops the context and returns $e$.}
    \Unwind_{\star}(e) &\eqdef e \\
    \Unwind_{\Gamma, x:A}(e) &\eqdef \Let p = \Swap^A_E(x, e) \In \delta\Paren{p, (e, x). \Paren{\Drop_A x; \Unwind_\Gamma(e)}}
\intertext{\item $\Gamma^+, e:E \vdash \Raise^A_\Gamma(e): A^-$ by induction on $A$, which drops the context and inhabits $A^-$ with $e$.}
    \Raise^{\Lmap{B}{C}}_\Gamma(e) &\eqdef \lambda b. \Raise^C_{b: B, \Gamma}(e) \\
    \Raise^{B \With C}_\Gamma(e) &\eqdef \langle \Raise^B_\Gamma(e), \Raise^C_\Gamma(e) \rangle \\
    \Raise^P_\Gamma(e) &\eqdef \Let e' = \Unwind_\Gamma(e) \In \iota_2 e'
\end{align*}
\end{itemize}
We then translate \AffLang derivations with two functions noted
$\llbracket-\rrbracket$ defined by mutual induction on derivations
of the two kinds of judgements:
\begin{itemize}
  \item \AffLang values of type $\Gamma \vdash A$ are translated
  to \LinLang values of type $\Gamma^+ \vdash A^+$.
\item \AffLang expressions of type $\Gamma \vdash A$ are translated
  to \LinLang expressions of type $\Gamma^+ \vdash \Down A^-$.
\end{itemize}

We add explicit coercions to treat positive values as expressions with
the notation $\Coerc(v)$. The monad strength for $T\Em$ can be observed
when translating $\Let$ bindings of expressions, where the context of
the continuation is dropped in case of an error.
{
\allowdisplaybreaks
\begin{align*}
  ax :{}& \llbracket x \vdash x:A \rrbracket \eqdef x \\
  \DropB :{}& \llbracket \vdash \DropB: \Lmap{A}{1} \rrbracket \eqdef \langle \Drop_A, () \rangle \\
  \Coerc :{}& \llbracket \Gamma \vdash \Coerc(v): P \rrbracket \eqdef \langle \iota_1 \llbracket v \rrbracket , \DropCtx_\Gamma \rangle \\
  \Move :{}& \llbracket \Gamma,x:A,\Delta \vdash \Move (x) \In t:C \rrbracket \eqdef \llbracket t \rrbracket \\
  \New :{}& \llbracket \vdash \New : \Lmap{1}{R} \rrbracket \eqdef \Let x = \New () \In \delta\Paren{x, r.\Paren{\iota_1 r}, i.\Paren{i; \iota_2 \NewFail}} \\
  \Let_v :{}& \llbracket \Gamma, \Delta \vdash \Let x=(v:A) \In t:B \rrbracket \eqdef \Let x = \llbracket v \rrbracket \In \llbracket t \rrbracket \\
  \Let_t :{}& \llbracket \Gamma, \Delta \vdash \Let x=(t:P) \In u:A \rrbracket \\
  &\eqdef \Let s = \pi_1\llbracket t \rrbracket  \In \delta\Paren{s, x. \llbracket u \rrbracket , e. \Raise^{A\With I}_\Gamma\Paren{e}} \\
  \RaiseB :{}& \llbracket \vdash \RaiseB : \Lmap{E}{A} \rrbracket \eqdef \langle \lambda e. \Raise^A_{\star}(e), () \rangle \\
  \Try :{}& \llbracket \Gamma, \Delta \vdash \Try x \Leftarrow (t:P) \In u \Unless e \Rightarrow u' : B \rrbracket \\
  &\eqdef \Let s = \pi_1\llbracket t \rrbracket  \In \delta(s, x. \llbracket u \rrbracket , e. \llbracket u' \rrbracket) \\
  \otimes_i :{}& \llbracket \Gamma, \Delta \vdash (v, w): A\otimes B \rrbracket \eqdef (\llbracket v \rrbracket , \llbracket w \rrbracket ) \\
  \otimes_e :{}& \llbracket \Gamma, \Delta, \Gamma' \vdash \delta(v, (x, y).t): C \rrbracket \eqdef \delta(\llbracket v \rrbracket , (x, y). \llbracket t \rrbracket ) \\
  1_i :{}& \llbracket \vdash (): 1 \rrbracket \eqdef () \\
  1_e :{}& \llbracket \Gamma, \Delta, \Gamma' \vdash \delta(v, ().t): C \rrbracket \eqdef \delta(\llbracket v \rrbracket , (). \llbracket t \rrbracket ) \\
  \oplus_{i1} :{}& \llbracket \Gamma \vdash \iota_1v: A\oplus B \rrbracket \eqdef \iota_1\llbracket v \rrbracket \\
  \oplus_{i2} :{}& \llbracket \Gamma \vdash \iota_2v: A\oplus B \rrbracket \eqdef \iota_2\llbracket v \rrbracket \\
  \oplus_e :{}& \llbracket \Gamma, \Delta, \Gamma' \vdash \delta(v, x.t, y.u): C \rrbracket \eqdef \delta(\llbracket v \rrbracket , x.\llbracket t \rrbracket, y.\llbracket u \rrbracket ) \\
  \LmapSym_i :{}& \llbracket \Gamma \vdash \lambda x.t: \Lmap{A}{B} \rrbracket \eqdef \langle \lambda x. \pi_1\llbracket t \rrbracket , \DropCtx_\Gamma \rangle \\
  \LmapSym_e :{}& \llbracket \Gamma, \Delta \vdash v w: B \rrbracket \eqdef \langle (\pi_1\llbracket v \rrbracket) \llbracket w \rrbracket, \DropCtx_{\Gamma, \Delta} \rangle \\
  \With_i :{}& \llbracket \Gamma \vdash \langle t, u \rangle: A\With B \rrbracket \eqdef \langle \langle \pi_1\llbracket t \rrbracket , \pi_1\llbracket u \rrbracket  \rangle, \DropCtx_\Gamma \rangle \\
  \With_{e1} :{}& \llbracket \Gamma \vdash \pi_1v: A \rrbracket \eqdef \langle \pi_1\pi_1\llbracket v \rrbracket, \DropCtx_\Gamma \rangle \\
  \With_{e2} :{}& \llbracket \Gamma \vdash \pi_2v: B \rrbracket \eqdef \langle \pi_2\pi_1\llbracket v \rrbracket, \DropCtx_\Gamma \rangle
\end{align*}}Note that expressions $t:P$ have a second component that is
immediately discarded, and they are evaluated eagerly. Hence, they are
essentially expressions of type $A \oplus E$. The operational
semantics of an affine value $v$ is that of $\llbracket v \rrbracket$,
and the operational semantics of an affine expression $t$ is that of
$\pi_1\llbracket t \rrbracket$.

\begin{proposition}
The translation defined above is well-defined: it preserves typing,
and it uses the exchange rule in the target only in the translation of
$\Move$, so the translation of the \OAffLang fragment indeed falls
into \OLinLang.
\end{proposition}

More details for the preservation of typing are given
in \cref{proof:translation}.

\begin{theorem}
For any expression $t$ of \AffLang with a purely-positive type without
resource $W$,
if $\langle \pi_1\llbracket
t \rrbracket \mid \star\mid l \rangle \rightsquigarrow^* \langle
v\mid \star\mid l' \rangle$ then $l'= \sigma(l)$ for some permutation
$\sigma$. Moreover, if the expression does not
contain ``$\Move$'', then $l'= l$.
\end{theorem}

\begin{proof}
Since purely-positive types without resources are translated into
themselves, one has $\Down W^- = (W \oplus E) \With I$, hence
$\pi_1\llbracket t \rrbracket$ is a linear expression of type
$W \oplus E$. By hypothesis $E$ is purely-positive without resource,
and so $W \oplus E$ is too. Hence, the resource-safety property
of \LinLang applies to $\pi_1\llbracket t \rrbracket$, as well as that
of \OLinLang if $t$ does not contain ``$\Move$''. Those are precisely
the properties we needed to prove.
\end{proof}

\section{A presheaf model of \OLinLang}
\label{sec:cat-model}

The ordered calculus \OLinLang shows some peculiarities compared to
usual formulations of ordered logic: it has a left arrow and no right
arrow, and a right let binding without left let binding. We find it illustrative to give a
simple concrete model showing that these peculiarities are natural and
arise by construction upon consideration of the strengths for the
allocation monad.

We interpret the calculus \OLinLang in the presheaf category
$\Ccat \eqdef \Set^{\BaseCat}$, with $\BaseCat$ the set of
lists of natural numbers considered as a discrete category. We note
$\Concat$ the non-symmetric monoidal product on $\BaseCat$
obtained by concatenating two lists of resources. Its Day
convolution~\cite{Day70} lifts it to a non-symmetric monoidal product in $\Ccat$ with a right closed and left closed structure, corresponding to the multiplicative fragment. We define
$R\in \Set^{\BaseCat}$ to be the indicator function for singleton
lists. As a presheaf category, $\Ccat$ has products and
coproducts. The initial algebra of $1 \oplus (R \otimes -)$
interpreting lists of resources is the terminal object $\top$.
Concretely, we have the following constructions in $\Ccat$:
\begin{itemize}
    \item The monoidal product $(A \otimes B)(l) \eqdef \sum_{(l_1,l_2) \in [R]^2 \wedge (l_1 \Concat l_2 = l)} A(l_1) \times B(l_2)$.
    \item The unit $1([]) \eqdef \{()\}, 1(l) \eqdef \emptyset$ is the indicator function for the empty list.
    \item The right arrow $(\Rmap{A}{B})(l_1) \eqdef \prod_{l_2 \in [R]} A(l_2) \to B(l_1 \Concat l_2)$.
    \item The left arrow $(\Lmap{A}{B})(l_2) \eqdef \prod_{l_1 \in [R]} A(l_1) \to B(l_1 \Concat l_2)$.
    \item The type of resources $R([n]) \eqdef \{n\}, R(l) \eqdef \emptyset$.
    \item The type of lists of resources $[R](l) \eqdef \{l\}$.
    \item The product $(A \With B)(l) \eqdef A(l) \times B(l)$.
    \item The coproduct $(A \oplus B)(l) \eqdef A(l) + B(l)$.
\end{itemize}

We exploit the fact that the canonical adjunction
\[
  F\eqdef (- \otimes [R]) \dashv G\eqdef \Rmap{[R]}{-}
\]
for the allocation monad is made of endofunctors such that
\begin{align}
  A \otimes FB \cong F(A \otimes B)\; \label{eq:adj-strength}
\end{align}
to interpret positive and negative terms in the same category: this allows us to adapt the model theory of \cite{CFM15} while avoiding the formalism of enriched categories.

We now have all the ingredients to interpret derivations of \OLinLang
(\cref{fig:typing}). In fact, we will interpret derivations of ordered
terms with resources from \cref{fig:ordered-rules} which extends the
typing rules of \OLinLang. Expressions typed with $ \Gamma;
[r_1 \cdots r_n]; \Delta \vdasho A$ are interpreted as (oblique)
morphisms in $\Ccat(F(\Gamma \otimes R^n \otimes \Delta),A)$,
following~\cite{CFM15} which we adapt to the ordered case
\ifbool{arxiv}{(see \cref{sec:cat-interp} for the complete definition of the interpretation)}{}.

Interpreting the left function and the right $\Let$ binding uses the
isomorphism (\ref{eq:adj-strength}) as a left strength. We cannot
interpret a right function or left $\Let$-binding for \OLinLang, as it
would require an isomorphism $FA \otimes B \cong F(A \otimes B)$
which swaps the ambient list of resources with $B$. This corresponds
to the re-ordering of resources by the incorrect term $t_r$
(\ref{def-incorrect-term}).

\begin{lemma}[Substitution lemma]\label{lem:cat-subst}
For all $\Theta'' \vdasho v:A$ and $\Theta,
x:A\CtxComp \Theta' \vdasho t:B$, one has $\llbracket t[v/x] \rrbracket = (id_{\Theta^+} \otimes \llbracket v \rrbracket \otimes
id_{\Theta'^+});\llbracket t \rrbracket$
\ifbool{arxiv}{ (see \cref{proof:cat-subst}).}{}
\end{lemma}

\begin{theorem}[Soundness of the interpretation]\label{thm:cat-sound}
For all $L,c,c'$ such that $L \vdasho c$ and $L \vdasho c'$, if
$c \rightsquigarrow c'$ then $\llbracket c \rrbracket = \llbracket
c' \rrbracket$\ifbool{arxiv}{ (see \cref{proof:cat-sound})}{}.
\end{theorem}

\mypar{Exceptions and destructors}
\label{sec:movable}

Exceptions need to be chosen among central objects in order to define
a strength for the exception monad. We ask for objects that commute
with all other objects $A$ in a coherent way. Those are the objects in
the \emph{Drinfeld centre} of $\Ccat$ (as defined e.g. in
\cite[§3]{Muger03}).

\begin{proposition}\label{thm:movable}
Given an object $E$, the three following properties are equivalent:
\begin{itemize}
  \item $E$ is in the Drinfeld centre of $\Ccat$.
  \item $E$ is resource-free: for all non-empty list $l$, $E(l) = \emptyset$.
  \item $E$ has a trivial destructor: there is a morphism $\Ccat(E, I)$.
\end{itemize}
\end{proposition}
\ifbool{arxiv}{See the proof in \cref{proof:movable}. }{}By induction, central
types in \OLinLang are resource-free,
therefore they belong to the Drinfeld centre of $\Ccat$.

\begin{proposition}
Every object in $\Ccat$ has a (unique) destructor, but $\Ccat$ is not
a model of affine logic in the sense of having an isomorphism
$I \cong \top$~(cf. \cite[§3.2]{Brauener1994}).
\end{proposition}

\begin{proof}
Recall that $[R]=\top$ is the terminal object. Destructors for $A$ are
objects in $(A,\delta)\in \Ccat/TI$; i.e. $\delta\in\Ccat(A,
TI) \cong \Ccat(A, [R] \multimap
[R]) \cong \Ccat(A \otimes \top, \top)$ and no other condition. By
the universal property of $\top$ there exists a unique such morphism.
On the other hand, only central objects have a trivial destructor
$\Ccat(A, I)$, which is not the case for $[R]=\top$.
\end{proof}

Given that any $\Gamma$ has a unique destructor, one has
$\Ccat/TI\cong\Ccat$. In particular, the monad $T\Em$ is left strong.

\begin{proposition}
Let $E$ a central object. The monad $T\Em$ where $\Em=(-\oplus E)$ has a
left strength $\Gamma\otimes T\Em A\rightarrow T\Em(\Gamma\otimes
A)$.\end{proposition}

\noindent Thus, this gives a model of an affine logic in the sense of the calculus with
destructors \OAffLang, but we do not have a model of affine logic in
the usual sense of semi-cartesian monoidal categories. Even if we
consider that the monoidal unit $I$ is indeed terminal in the Kleisli
category of $T$, the destructors are not thunkable for $T$, a common
requirement for morphisms interpreting structural rules in premonoidal
categories (see e.g. Führmann~\cite{Fuhrmann99}). This is another way
of observing the fact that weakening performs a side effect.

We might wonder how we could extend this model to ones of \LinLang and \AffLang.
This amounts to somehow regaining the symmetry of the monoidal product
to interpret the exchange rule. One way to do that would be to
quotient our base category $\BaseCat$ to obtain multisets. However,
$T$ would become commutative and so $\Ccat_{/TI}$ would be symmetric;
intuitively the model does not observe the order of destruction.
Instead, adding permutations as isomorphisms in $\BaseCat$ would
result in a presheaf model on a setoid in which $T$ is not
commutative; intuitively the model tracks in the interpretation of
terms their action on the free-list of resources. As we cannot re-use
our trick with endofunctors to avoid the presheaf-enriched formalism,
we leave this as future work.

\section{Discussion and perspectives}
\label{sec:discussion}

After this formal development, we find it useful to place our
results in the context of the theory and practice of programming
languages.

\mypar{Integrating linear types and error handling} We modelled a
global allocator with a linear state monad to study notions of
resource-safety for linear calculi: for instance, every allocated
resource in \LinLang and \AffLang is freed exactly once, even in case
of errors during the execution of the program. This result strengthens
when the exchange rule (\ref{eq:exchange}) is not used: in \OLinLang
and \OAffLang, the resources are released in a LIFO order. A
relationship between ordered logic and stack-like allocation was
proposed
previously~\cite{Polakow2001PhD,Polakow2001,Walker2005a,Pfenning2009}.
In this context, our results suggest to consider a variant of ordered
logic that only has one arrow type and whose composition (let) is
restricted to the opposite side of abstraction ($\lambda$)---where in
particular $\lambda$ cannot express let. The modelling of the exchange
rule with a side-effecting ``move'' operation is also novel.

Conceptually close to stack allocation, LISP's higher-order combinator
\linebreak \verb`unwind-protect` that executes a clean-up function
upon normal or exceptional return of its argument (like its variants
found in modern academic programming languages such as OCaml and
Haskell) has the same constraint of deallocation in LIFO order.

``Linear types,'' on the other hand, promised to let us manipulate
resources as first-class values (see~\cite{Baker94,Baker95} among
others). However, extensive practical experiments with linear types
stumbled (among other) on the problematic interaction of linearity
with error handling, another important aspect with resources that are
acquired in a program. For instance, Cyclone, a source of inspiration
for the Rust language, permitted resource leaks in the name
of \emph{``flexibility and usability''}~\cite[§3.5]{Swamy2006}
(necessitating back-up collection mechanisms). The practical
experiment of Tov and Pucella~\cite{Tov11,Tov2012}, another milestone
in exploring practical aspects of linear type systems, also mentioned
the issue of combining linearity and control effects, which became the
motivation for their \emph{``practical''} linear type system to be
affine.

Tov and Pucella~\cite{Tov11Exn}, separately, seeking to lift the
limitation of the affine system for manipulating resources as
first-class values, proposed a type system that offers both linearity
and control effects (e.g. checked exceptions), by typing effects in
addition to linearity. Essentially, effects are constrained when
linear variables are in scope (or conversely), matching the
restrictive approach we described in \cref{sec:strength-wrt}
corresponding to the notion of strength with respect to an affine or
an unrestricted resource modality (\ref{eq:strength-relative}). It
should be noted that this system~\cite{Tov11Exn}, unlike the previous
experiment~\cite{Tov11}, is mainly justified by its mathematical
properties; how useful such a system is, in practice, remains
unclear.\footnote{As Tov themselves explains: \emph{``One question that
remains, however, concerns the pragmatics of checked exceptions in a
higher-order language such as Alms, where latent exception effects are
likely to appear on many function arrows. Weighing the cost against
the benefit, I have decided that adding linear types to Alms is not
worth the complexity of a programmer-visible effect
system.''}~\cite[p. 238]{Tov2012}}

Lastly, \emph{affine session
types}~\cite{Mostrous2018} do mix linearity and cancellation in a sense
similar to ours. They guarantee a safety property (absence of
deadlock) after an exception (cancellation) arises, by communicating
cancellation across channels.

Our analysis from the introduction applies equally to explicit error
handling with an error monad, as illustrated with the language
$\LinLang$. Indeed, manual error handling encounters similar obstacles
arising from the lack of unrestricted monadic strength for the error
monad. Concretely, for each location where one would like to use a
monadic binding (\ref{eq:monadic-bind}), one needs to implement by
hand a repetitive description of whichever clean-up functions should
be called in case of error propagation---a well-known and tedious
problem in C, which gave rise to specific programming patterns
involving
\verb`goto`\ifbool{arxiv}{ (see \cref{sec:res-prog})}{}.

By 2012, these limitations of linear types with respect to the
handling of errors and exceptions, arising from practical
considerations, might seem well-established in the literature.
However, more recent works on linear type systems purporting to
implement practical first-class resources in academic programming
languages~\cite{Bernardy2018,Orchard2019} do not explore nor mention
these obstacles. In the case of \cite{Bernardy2018}, a section
describing limitations arising from Haskell exceptions indeed appears
in an earlier version with a different title~\cite{Bernardy2017}, but
is absent from the published peer-reviewed version.

With the resource CBPV, we set out to model some aspects
of C++ resources with destructors and move operations.
Compared to the works cited previously, C++ realised, much
earlier~\cite{Stroustrup1993History}, a shift in viewpoint whereby
clean-up functions are deduced from the types, whilst it also later
pioneered ``move semantics''~\cite{Hinnant2002} for treating resources
as first-class values. As we have seen, interpreting resource types in
the slice category over some $TI$, i.e.\ as types provided with
some chosen effectful weakening map $A\rightarrow TI$, suffices to
provide a notion of strength for the exception (or error) monad.
Concretely this means that we can both program with an error monad,
and give a meaning to exceptions as a side-effect. One surprising
aspect of the resource CBPV is that it combines facets
of all three of affine, linear and ordered logics:
\begin{center}
{\renewcommand{\arraystretch}{1.2}
\begin{tabular}{cl}
\emph{affine} &in terms of types and available control effects\tabularnewline
\hline
\emph{linear} &in terms of first-class values and resource-safety properties\tabularnewline
\hline
\emph{ordered} &in terms of proof-relevant semantics and type isomorphisms\tabularnewline
\end{tabular}
}
\end{center}
Given that the logical expressiveness is that of affine logic
(including the exchange rule), without sacrificing the linear
character of values (which are correctly released), we do not have to
choose between linearity and control effects. This perspective
suggests that it should be possible to adapt practical works on affine
typing (in the lineage of \cite{Tov11}) in order to integrate
first-class resources in functional programming
languages\ifbool{doubleblind}{}{~\cite{Munch18RePo}}.

\mypar{A model of C++/Rust-style resource management and its interpretation}The resource CBPV is a very idealised model of
C++/Rust-style resource management situated at the intersection of
linear logic and the theory of effects. Various models of ownership in
Rust---more practical and less idealised---have been developed to prove safety and
functional properties of Rust programs, such as with separation
logic~\cite{Jung2018}, a dedicated functional logic~\cite{Denis2022},
or by translation into a purely functional language~\cite{Ho2022}.
These recent works, explicitly, do not model exceptions with
destructor calls, and they do not develop a specific understanding of
destructors in relationship with linearity.

To our knowledge, our model is the first to reproduce various
phenomena seen with C++/Rust's resource management such as: the
propagation of errors involving the release of resources in scope as
part of monadic binding; destructors having to never fail; resources
being first-class values that can be passed, returned, or stored in
algebraic data types and closures, and for which ``moving'' resources
(identified with an exchange rule in logic, albeit performing a
side-effect) is responsible for altering the order of destruction; and
indeed closures themselves being resources. In fact, in the presence
of both resource modalities $\oc$ and $\Dm$, different positive types
of closures, derived from $\LmapSym$, coexist ($\oc(A\LmapSym B)$ and
$\Dm(A\LmapSym B)$), distinguished by the restrictions on the typing
context in accordance with the resource modality (e.g. copyable or
resource-like)---a distinction between several types of closures that
can also be seen in C++ and Rust.

One very interesting aspect of the resource CBPV is that
exchange and weakening are interpreted by morphisms that perform
effects. For instance, $A\otimes B$ and $B\otimes A$ are not
contextually isomorphic in the language with destructors because they
can be distinguished observationally from the order in which the
destructors of $A$ and $B$ are executed. So we can have a logical
equivalence in the form of inverse morphisms:
\begin{equation}
  A\otimes B \rightleftarrows B\otimes A\label{eq:equiv-tensor}
\end{equation}
without the two types being isomorphic. This is because
a \emph{contextual} notion of isomorphism (allowing to substitute
equals by equals) also requires the two inverse
maps~(\ref{eq:equiv-tensor}) to be pure~\cite{Levy2017}. This
radically departs from symmetric premonoidal
categories~\cite{Power1997}.

We usually see the Curry-Howard correspondence at work when algebraic
and logical structures inspire new programming language features.
Could the concept of destructors have appeared out of theoretical
consideration or practical experiments starting from linear logic? As
it turns out, the interpretation of resource types with destructors as
ordered logic formulae almost arose on several occasions, such as when
Baker suggested that C++ constructors and destructors could fit within
his linear language, in rarely-cited essays that anticipated C++ move
semantics~\cite{Baker94,Baker95}, and when Gan, Tov and Morrisett
approached substructural types using type classes \verb`dup` and
\verb`drop`~\cite{Gan2014}. But we find it remarkable that the
Curry-Howard correspondence actually worked this time in the converse
direction, with C++ destructors and move semantics arising from
practical consideration over more than 20
years~\cite{KoenigStroustrup90,Stroustrup1993History,Hinnant2002}, and
with the use of the slice category in this context being prompted by
empirically-observed phenomena with C++
resources~\cite{CombetteMunch18}.

\mypar{Conclusion and future work}
Our hope is that this newfound understanding of resources in
programming will be inspire improvements to programming languages and
designs for better ones. It could also help proving more properties of
programs with formal methods, such as methods based on translations
into pure functional programs. Indeed, the functional correctness of
some Rust programs can depend on the resource-safety properties
enforced by the language, for instance when they use the typestate
programming pattern or when they are made fault-tolerant.

The Curry-Howard correspondence is sometimes stated as a more
technical result relating logic, categorical structures and
programming language models, as in the Curry-Howard-Lambek
correspondence for linear CBPV~\cite{CFM15} which the
present work is based on. Though not essential to the results of this
paper, it is now clear that extending the Curry-Howard-Lambek
correspondence to ordered CBPV would be useful as a
meta-theory to study resources. Removing the symmetry in monoidal
categories and the exchange rule in logic can create technical
issues---we have made some progress already by understanding that
$\lambda$ and let bindings work on opposite sides, and that being
merely left-closed or right-closed but not both is already interesting
and useful.

Lastly, much remains to be understood in an idealised and principled
manner within this framework, the most obvious ones being:

\begin{itemize}
\item\textbf{Copyable types } Linear languages, including
with C++ and Rust, mix in practice resource types and unconstrained
types. The interaction of the resource modality arising from the slice
construction with the resource modality $\oc$ which controls freely
copyable and erasable types, should therefore be explored.
A general mechanism of polarities should provide foundations for
kind-based linear type systems~\cite{Tov11} and extend them to
computationally-relevant kinds in the style of Rust's special traits
(e.g. \verb`Copy`, \verb`Drop`). Interesting questions arise from the practical
necessity of kind polymorphism in such languages.

\item\textbf{Borrowing } We have modelled resources that can only be
allocated and deallocated, which lets us observe the effect of
deallocations on the final state of the program. More questions arise
when resources can be used between their allocation and their
deallocation. Operations that do not consume the resource can be
implemented by returning it~\cite{Baker94,Baker95}.
Borrowing~\cite{Grossman2002,Fluet2006}, inspired by
regions~\cite{Tofte1997}, and refined into unique borrowing in Rust,
was introduced as a more usable and expressive alternative to this
linear threading of values. As we already mentioned, various works
account for borrowing using different
approaches~\cite{Jung2018,Denis2022,Ho2022}, but it remains to be seen
whether and how borrowing may be derived in a principled way from a
concept of linearity (ownership) arising from destructors.

\end{itemize}

\subsection*{Acknowledgements}
We thank the anonymous reviewers for their detailed and thoughtful
comments on earlier versions of this article.

\printbibliography

\ifbool{arxiv}{}{\end{document}}

\appendix

\section{Context: resource management in programming}
\label{sec:res-prog}

Programmers must sometimes deal with \emph{resources}: values denoting
some memory allocations, files, locks, etc.---in general reifying some
knowledge about the state of the world---that must be disposed of at
specific points in the program to ensure its proper functioning. For
instance, some task might operate on a file descriptor, obtained and
released through operating system calls, as illustrated with the
following pseudo-code:

\begin{lstlisting}[caption={Opening and closing a file}]
let f = open_file("record.txt");
append(f, "starting tasks");
run_tasks(f);
close_file(f);
\end{lstlisting}

It is generally a good idea to close files in a timely manner: for
instance, depending on the operating system and its command, the
number of files that the program is allowed to keep opened
concurrently can be fairly low.
Another common phenomenon affecting resources is that their
acquisition can fail, which actually complicates the problem of
correctly disposing of resources: it must be carefully done in case of
failure of later resource acquisitions. This is where C programmers
resort to using \verb`goto`---one of the only situations where it is
not only allowed but idiomatic; in pseudo-code:
\begin{lstlisting}[caption={Handling errors with goto}]
  let f = open_file("record.txt");
  if !f return;
  append(f, "starting tasks");
  let g = open_file("output.txt");
  if !g goto err;
  run_tasks(f, g);
  close_file(g);
err:
  close_file(f);
\end{lstlisting}
Here, \verb`f` is closed whether opening \verb`g` succeeds or not.

Managing resources, especially explicitly as in C, comes with specific
concerns:
\begin{enumerate}
    \item \label{i1}disposing of the resource in a timely fashion in all
    the code paths (avoiding leaks, as we have seen),
    \item \label{i2}keeping track of who is responsible for disposing of
    the resource (to avoid for instance that the resource is disposed of twice),
    \item \label{i3}making sure that a resource is never used after its
    disposal (\emph{use-after-free}).
\end{enumerate}

In simple situations like the previous one, and in languages that
support it, the resources can be managed automatically with
higher-order scoped combinators derived from
LISP's \verb`unwind-protect`:
\begin{lstlisting}[caption={Handling errors with higher-order scoped combinators}]
with_file("record.txt", LAMBDA f.
  append(f, "starting tasks");
  with_file("output.txt", LAMBDA g.
    run_tasks(f, g)
  );
);
\end{lstlisting}
In words, the combinator \verb`with_file` opens the file and passes
the file descriptor to its argument. It closes the file upon normal or
exceptional return, and returns the same value that is returned by the
functional argument.

Now, this approach with scoped combinators comes with several
limitations. Firstly, they force resources to be disposed of in a
strict LIFO order. They do not model situations where the duty of
clean-up is transferred to the caller or to a callee, nor where the
resource is stored in a data structure. This limits expressiveness but
also compositionality, the ability to reorganise the program in
smaller and/or more general parts. In effect, scoped combinators
solve \eqref{i1} and \eqref{i2} above albeit in a heavy-handed way.

The second limitation has to do with the transient nature of the value
they pass to their functional argument: this argument is indeed
meaningful for the duration of the scope only, since using the value
after the scope ends (by sneaking it inside the return value, or by
storing it inside some mutable data structure) is an attempt to use
the resource after its disposal \eqref{i3}. Yet, keeping track of the
value can be far from obvious, especially when programming using other
(e.g.\ monadic) combinators.

We focused in this paper on using a notion of
\emph{ownership},
generally understood to be related to linearity, to lift the first
limitation using. We elaborated more precisely a formal link between a
notion of ownership present in the C++ and Rust programming languages,
that has not seen much theoretical study, and linearity.
We did not attempt to model the second limitation, which is already
the subject of much study involving other theoretical tools.

\mypar{C++ destructors}

Independently from linear types, modern systems programming languages
such as C++ and Rust have since the 1980's evolved
automatic resource-management features that not only combine well with
error management and control effects, but in which error handling is
an essential part---indeed resource management requires to deal with
the correct disposal of resources, but also the handling of possible
failure of their allocation. Furthermore, links with linear logic were
suggested as early as~\cite{Baker94}, but remained elusive.

Each type $A$ comes equipped with a \emph{destructor}, a clean-up
action that removes a value of $A$. It is called when such variable
goes out of scope, because of an explicit return or an exception.
Destructors provide a notion of ownership, tying resources to
variables lifetime. Local variables hence yields a LIFO ordering for
handling resources. The previous example would be written like this in
C++:

\begin{lstlisting}[caption={Handling errors with destructors}]
File f{"record.txt"};
append(f, "starting tasks");
File g{"output.txt"};
run_tasks(f, g);
// g then f are then closed by their destructors
\end{lstlisting}

Sometimes, the control flow of resources becomes more complex. For
example, in the following example, we transfer the file ownership
to \verb`run_task`, in the case where tasks have been initialised. It
then becomes difficult to track the places responsible to
call \verb`close_file`, leading to potential leaks or errors. Also,
this use case does not fit in the pattern of higher-order combinators,
which would always close the file at the end of the scope.

\begin{lstlisting}[caption={Pseudo-code for manual conditional transfer of ownership}]
let f = open_file("log.txt");
if !f return;
if init_tasks() == ok:
    append(f, "starting tasks");
    run_tasks(f);
else:
    append(f, "failure to init tasks");
    close_file(f);
\end{lstlisting}

To implement the transfer of ownership, C++ initially experimented
with smart pointers such as the now-deprecated \verb`auto_ptr`.
So-called ``move semantics'' was introduced in C++11 to fix the
defects of \verb`auto_ptr` in representing ownership and the excessive
amounts of deep copies encouraged by the language. So-called ``move
constructors'' and ``move assignment operators'', associated to a
notion of ``rvalue reference'', enabled the definition of ``unique
pointers'' and a ``move'' operation to transfer ownership of values,
as illustrated below:

\begin{lstlisting}[caption={C++ pseudo-code using a file to record operations}]
File f{"log.txt"};
if init_tasks() == ok {
    append(f, "starting tasks");
    run_tasks(std::move(f));
} else {
    append(f, "failure to init tasks");
}
\end{lstlisting}

\noindent In effect, \verb`std::move` is responsible for selecting the move
constructor of \verb`File` for its argument; the move constructor is
then responsible for implementing the move operation which consists
(most of the time) in a shallow copy into the target, followed by the
nullification of the source one (or at least its replacement with some
valid, default value). In this way, it is guaranteed that a single
copy of the resource exists at any time, and the destructor
of \verb`f` is not called (or more precisely, is called on a null or
default value, which does nothing). This notion of ownership in C++
can be described as ensuring that the owner (responible for calling
the destructor) is unique.

In order to implement first-class resources, C++11 introduced language
and standard library features making it convenient to define resource
datatypes, such as the rules for the automatic derivation of move
constructors (which, together with similar rules for destructors, are
responsible for instance for a struct of resources to behave as a
resource in the obvious manner). In actual usage (see e.g. the ``rule of zero''), these are very close
to defining separate kinds of movable vs. copyable types.

\allowdisplaybreaks

\section{Details: basic properties}

\mypar{Proof of $\Swap$ and definition of $\Swap$ inverse}
\label{proof:swap}

We recall the expressions $\Swap^A_W$:
\begin{align*}
\Swap^A_I &\eqdef \lambda p. \delta(p, (a, i). \delta(i, ().(\star, a))) \\
 \\
\Swap^A_{W_1 \oplus W_2} &\eqdef \lambda p. \delta(p, (a, w). \delta(w, \\
& w_1. \Let p_1 = \Swap^A_{W_1} (a, w_1) \In \delta(p_1, (w_1, a). (\iota_1 w_1, a)), \\
& w_2. \Let p_2 = \Swap^A_{W_2} (a, w_2) \In \delta(p_2, (w_2, a). (\iota_2 w_2, a)))) \\
 \\
\Swap^A_{W_1 \otimes W_2} &\eqdef \lambda p. \delta(p, (a, w). \delta(w, (w_1, w_2). \\
& \Let p_1 = \Swap^{A \otimes W_1}_{W_2} ((a, w_1), w_2) \In \delta(p_1, (w_2, p). \delta(p, (a, w_1). \\
& \Let p_2 = \Swap^{W_2 \otimes A}_{W_1} ((w_2, a), w_1) \In \delta(p_2, (w_1, p). \delta(p, (w_2, a). ((w_1, w_2), a)))))))
\end{align*}

We define similarly the expressions $\Swap'^A_W: \Lmap{W \otimes A}{A \otimes W}$:
\begin{align*}
\Swap'^A_I &\eqdef \lambda p. \delta(p, (i, a). \delta(i, ().(a, \star))) \\
 \\
\Swap'^A_{W_1 \oplus W_2} &\eqdef \lambda p. \delta(p, (w, a). \delta(w, \\
& w_1. \Let p_1 = \Swap'^A_{W_1} (w_1, a) \In \delta(p_1, (a, w_1). (a, \iota_1 w_1)), \\
& w_2. \Let p_2 = \Swap'^A_{W_2} (w_2, a) \In \delta(p_2, (a, w_2). (a, \iota_2 w_2)))) \\
 \\
\Swap'^A_{W_1 \otimes W_2} &\eqdef \lambda p. \delta(p, (w, a). \delta(w, (w_1, w_2). \\
& \Let p_1 = \Swap'^{A \otimes W_1}_{W_2} (w_1, (w_2, a)) \In \delta(p_1, (p, w_1). \delta(p, (w_2, a). \\
& \Let p_2 = \Swap'^{W_2 \otimes A}_{W_1} (w_2, (a, w_1)) \In \delta(p_2, (p, w_2). \delta(p, (a, w_1). (a, (w_1, w_2))))))))
\end{align*}

Those terms are well-typed: variables are used exactly once and in order. Intro and elimination rules for positive types do not have ordering constraints. Those from function application and let bindings are respected: $\Swap$ arguments use right-most variables. Informally, those terms implement the following operations on stacks, with permutations restricted to the two right-most types:
\begin{itemize}
  \item $\Swap^A_I \eqdef A I \to A \to I A$.
  \item $\Swap^A_{W_1 \oplus W_2} \eqdef A (W_1 \oplus W_2) \to A W_i \to W_i A \to (W_1 \oplus W_2) A$.
  \item $\Swap^A_{W_1 W_2} \eqdef A (W_1 W_2) \to (A W_1) W_2 \to W_2 (A W_1) \to (W_2 A) W_1 \to W_1 (W_2 A) \to (W_1 W_2) A$.
  \item $\Swap_A^I \eqdef I A \to A \to A I$.
  \item $\Swap_A^{W_1 \oplus W_2} \eqdef (W_1 \oplus W_2) A \to W_i A \to A W_i \to A (W_1 \oplus W_2)$.
  \item $\Swap_A^{W_1 W_2} \eqdef (W_1 W_2) A \to W_1 (W_2 A) \to (W_2 A) W_1 \to W_2 (A W_1) \to (A W_1) W_2 \to A (W_1 W_2)$.
\end{itemize}

\mypar{Proof of determinism (\cref{prop:determinism})}
\label{proof:determinism}

This is shown by case analysis, first on the command polarity:

For negative commands, all expressions are mutually exclusive except the two ``$\New$'' cases, for which the lists are mutually exclusive (empty or not).

For positive commands, all expressions are mutually exclusive once we analyse $v$ in $\langle v\mid (x^+.t) \cdotp s\mid l \rangle^+ \rightsquigarrow \langle t[v/x]\mid s\mid l \rangle^+$: either the value is a variable or positive, in which case no other rules apply, or the value is negative and may reduce only in negative commands.

\mypar{Proof of subject reduction (\cref{thm:subject-reduction})}
\label{proof:subject-reduction}

Show that $\forall c1:A, \forall c2, c1 \rightsquigarrow
c1 \Rightarrow c2:A$ by case analysis on the reduction step:
\begin{itemize}
    \item $\langle (\Let x^+=t \In u)^\varepsilon\mid s\mid l \rangle^\varepsilon \rightsquigarrow \langle t\mid (x^+.u)^\varepsilon \cdotp s\mid l \rangle^+$: by the hypothesis $\langle (\Let x^+=t \In u)^\varepsilon\mid s\mid l \rangle^\varepsilon: A$, we get $\vdashp t:B_+, x:B \vdashp u:C_\varepsilon$ and $s:C \vdashp A$. Hence, we have $(x^+.u)^\varepsilon \cdotp s: B \vdashp A$ and so $\langle t\mid (x^+.u)^\varepsilon \cdotp s\mid l \rangle^+: A$.

    \item $\langle (\Let x^-=v \In t)^\varepsilon\mid s\mid l \rangle^\varepsilon \rightsquigarrow \langle t[v/x]\mid s\mid l \rangle^\varepsilon$: by hypothesis, we get $\vdashp v:B_-, x:B \vdashp t:C_\varepsilon$ and $s:C \vdashp A$. By SL, we have $\vdashp t[v/x]: C$ and so $\langle t[v/x]\mid s\mid l \rangle^\varepsilon: A$.

    \item $\langle (v w)^\varepsilon\mid s\mid l \rangle^\varepsilon \rightsquigarrow \langle v\mid w^\varepsilon\cdotp s\mid l \rangle^-$: by hypothesis, we get $\vdashp v: \Lmap{B}{C_\varepsilon}, \vdashp w: B$ and $s: C \vdashp A$. Hence, we have $w^\varepsilon \cdotp s: \Lmap{B}{C} \vdashp A$ and so $\langle v\mid w^\varepsilon\cdotp s\mid l \rangle^-: A$.

    \item $\langle (\pi_i v)^\varepsilon\mid s\mid l \rangle^\varepsilon \rightsquigarrow \langle v\mid \pi_i^\varepsilon\cdotp s\mid l \rangle^-$: by hypothesis, we get $\vdashp v: B_1 \With B_2$ and $s: B_{i\varepsilon} \vdashp A$. Hence, we have $\pi^\varepsilon_i \cdotp s: B_1 \With B_2 \vdashp A$ and so $\langle v\mid \pi_i^\varepsilon\cdotp s\mid l \rangle^-: A$.

    \item $\langle v\mid (x^+.t)^\varepsilon \cdotp s\mid l \rangle^+ \rightsquigarrow \langle t[v/x]\mid s\mid l \rangle^\varepsilon$: by hypothesis, we get $\vdashp v: B_+, x: B \vdashp t: C$ and $s: C \vdashp A$. By SL, we have $\vdashp t[v/x]: C$ and so $\langle t[v/x]\mid s\mid l \rangle^\varepsilon: A$.

    \item $\langle \lambda x.t\mid v^\varepsilon \cdotp s\mid l \rangle^- \rightsquigarrow \langle t[v/x]\mid s\mid l \rangle^\varepsilon$: by hypothesis, we get $x: B \vdashp t: C_\varepsilon, \vdashp v: B$ and $s: C \vdashp A$. By SL, we have $\vdashp t[v/x]: C$ and so $\langle t[v/x]\mid s\mid l \rangle^\varepsilon: A$.

    \item $\langle \langle t_1, t_2 \rangle\mid \pi_i^\varepsilon \cdotp s\mid l \rangle^- \rightsquigarrow \langle t_i\mid s\mid l \rangle^\varepsilon$: by hypothesis, we get $\vdashp t_1: B_1, \vdashp t_2: B_2$ and $s: B_{i\varepsilon} \vdashp A$. Hence, we have $\langle t_i\mid s\mid l \rangle^\varepsilon: A$.

    \item $\langle \delta ((v,w), (x,y).t)^\varepsilon\mid s\mid l \rangle^\varepsilon \rightsquigarrow \langle t[v/x, w/y]\mid s\mid l \rangle^\varepsilon$: by hypothesis, we get $\vdashp v: B, \vdashp w: C, x:B, y:C \vdashp t: D_\varepsilon$ and $s: D \vdashp A$. By SL, we have $\vdashp (t[v/x])[w/y]: D$. Because $x$ nor $y$ do not occur free in $v$ nor $w$, we have $(t[v/x])[w/y] = t[v/x, w/y]$ and so $\langle t[v/x, w/y]\mid s\mid l \rangle^\varepsilon: A$.

    \item $\langle \delta ((), ().t)^\varepsilon\mid s\mid l \rangle^\varepsilon \rightsquigarrow \langle t\mid s\mid l \rangle^\varepsilon$: by hypothesis, we get $\vdashp t: B_\varepsilon$ and $s: B \vdashp A$, hence we have $\langle t\mid s\mid l \rangle^\varepsilon: A$.

    \item $\langle \delta (\iota_iv, x_1.t_1, x_2.t_2)^\varepsilon\mid s\mid l \rangle^\varepsilon \rightsquigarrow \langle t_i[v/x_i]\mid s\mid l \rangle^\varepsilon$: by hypothesis, we get $\vdashp v: B_i, x_1: B_1 \vdashp t_1: C_\varepsilon, x_2: B_2 \vdashp t_2: C$ and $s: C \vdashp A$. By SL, we have $\vdashp t_i[v/x_i]: C$ and so $\langle t_i[v/x_i]\mid s\mid l \rangle^\varepsilon: A$.

    \item $\langle \New\mid ()\cdotp s\mid r_n\cons l \rangle^- \rightsquigarrow \langle \iota_1r_n\mid s\mid l \rangle^+$: by hypothesis, we have $s: R \oplus 1 \vdashp A$. We always have $\vdashp \iota_1r_n: R \oplus 1$, and so $\langle \iota_1r_n\mid s\mid l \rangle^+:A$.

    \item $\langle \New\mid ()\cdotp s\mid \Nil \rangle^- \rightsquigarrow \langle \iota_2()\mid s\mid \Nil \rangle^+$: by hypothesis, we have $s: R \oplus 1 \vdashp A$. We always have $\vdashp \iota_2(): R \oplus 1$, and so $\langle \iota_1r_n\mid s\mid l \rangle^+:A$.

    \item $\langle \Delete\mid r_n\cdotp s\mid l \rangle^- \rightsquigarrow \langle ()\mid s\mid r_n\cons l \rangle^+$: by hypothesis, we have $s: 1 \vdash A$. We always have $\vdashp (): 1$ and so $\langle ()\mid s\mid r_n\cons l \rangle^+$.
\end{itemize}

The same proof goes through if we consider $\vdashp$ without structural rules.

\mypar{Proof of progress (\cref{thm:progress})}
\label{proof:progress}

Show that any well-typed command $\langle t \mid s \mid l \rangle^\varepsilon:A$ reduces iff it is not of the shape $\langle v_t\mid \star\mid l \rangle^\varepsilon$ by case analysis on the judgement $\vdashp t:B_\varepsilon$:

\noindent
If $t$ is a final value, we do a case analysis on the judgement $s: B \vdashp A$: if $s=\star$, the context is of the shape $\langle v_t\mid \star\mid l \rangle^\varepsilon$ and matches no reduction steps. Otherwise, the stack shape is determined by the type of $t$:
\begin{itemize}
    \item If $\varpi(B)=+$, then $s=(x^+.u)^\varepsilon\cdotp s'$ and $\langle v_t\mid (x^+.u)^\varepsilon\cdotp s'\mid l \rangle^+ \rightsquigarrow \langle u[v_t/x]\mid s'\mid l \rangle^\varepsilon$.
    \item If $t=\langle t_1, t_2 \rangle$, then $s=\pi_i^\varepsilon\cdotp s'$ and $\langle \langle t_1, t_2 \rangle\mid \pi_i^\varepsilon\cdotp s'\mid l \rangle^- \rightsquigarrow \langle t_i\mid s'\mid l \rangle^\varepsilon$.
    \item If $t=\lambda x.t'$, then $s=v^\varepsilon\cdotp s'$ and $\langle \lambda x.t'\mid v^\varepsilon\cdotp s'\mid l \rangle^- \rightsquigarrow \langle t[v/x]\mid s'\mid l \rangle^\varepsilon$.
    \item If $t=\New$ and $l=r_n\cons l'$, then $s=()\cdotp s'$ and $\langle \New\mid ()\cdotp s'\mid r_n\cons l' \rangle^- \rightsquigarrow \langle \iota_1r_n\mid s'\mid l' \rangle^+$.
    \item If $t=\New$ and $l=\star$, then $s=()\cdotp s'$ and $\langle \New\mid ()\cdotp s'\mid \star \rangle^- \rightsquigarrow \langle \iota_2()\mid s'\mid \star \rangle^+$.
    \item If $t=\Delete$ so $s=r_n\cdotp s'$ and $\langle \Delete\mid r_n\cdotp s'\mid l \rangle^- \rightsquigarrow \langle ()\mid s'\mid r_n\cons l \rangle^+$.
\end{itemize}

\noindent
Otherwise:
\begin{itemize}
    \item If $t=\delta ((v,w), (x,y).t')^\varepsilon$, then $\langle \delta ((v,w), (x,y).t')^\varepsilon\mid s\mid l \rangle^\varepsilon \rightsquigarrow \langle t'[v/x, w/y]\mid s\mid l \rangle^\varepsilon$.
    \item If $t=\delta ((), ().t')^\varepsilon$, then $\langle \delta ((), ().t')^\varepsilon\mid s\mid l \rangle^\varepsilon \rightsquigarrow \langle t'\mid s\mid l \rangle^\varepsilon$.
    \item If $t=\delta (\iota_iv, x_1.t_1, x_2.t_2)^\varepsilon$, then $\langle \delta (\iota_iv, x_1.t_1, x_2.t_2)^\varepsilon\mid s\mid l \rangle^\varepsilon \rightsquigarrow \langle t_i[x_i/v]\mid s\mid l \rangle^\varepsilon$.
    \item If $t=(\Let x^-=v \In t')^\varepsilon$, then $\langle (\Let x^-=v \In t')^\varepsilon\mid s\mid l \rangle^\varepsilon \rightsquigarrow \langle t'[v/x]\mid s\mid l \rangle^\varepsilon$.
    \item If $t=(\Let x^+=t' \In u)^\varepsilon$, then $\langle (\Let x^+=t' \In u)^\varepsilon\mid s\mid l \rangle^\varepsilon \rightsquigarrow \langle t'\mid (x^+.u)^\varepsilon \cdotp s\mid l \rangle^+$.
    \item If $t=(v w)^\varepsilon$, then $\langle (v w)^\varepsilon\mid s\mid l \rangle^\varepsilon \rightsquigarrow \langle v\mid w^\varepsilon\cdotp s\mid l \rangle^-$.
    \item If $t=(\pi_i v)^\varepsilon$, then $\langle (\pi_i v)^\varepsilon\mid s\mid l \rangle^\varepsilon \rightsquigarrow \langle v\mid \pi_i^\varepsilon\cdotp s\mid l \rangle^-$.
\end{itemize}

The same proof goes through if we consider $\vdashp$ without
structural rules.

\section{Details: resource-safety properties}

\mypar{Proof that $\vdash$ implies resource-free $\vdasho$ (\cref{lem:vdasho})}
\label{proof:vdasho}

\begin{proof}
By induction on the derivation of $t$, we apply the corresponding rule in $\vdasho$: contexts of premises are always well-formed due to the absence of resources. Rules to shift between the contexts $\Gamma, x; []; \Delta$ and $\Gamma; []; x, \Delta$ are used as needed, in particular at the end to obtain $\Gamma;; \vdash t$.
\end{proof}

\mypar{Proof of left and right substitution lemmas for ordered expressions (\cref{lem:ord-sl})}
\label{proof:ord-sl}

Show that (Left SL) if $;L_t; \vdasho t$ and $\Gamma, x;
L_u; \Delta \vdasho u$, then $\Gamma; L_t \Concat L_u; \Delta \vdasho
u[t/x]$. Moreover (Right SL), if $;L_t; \vdasho t$ and $\Gamma; L_u;
x, \Delta \vdasho u$, then $\Gamma; L_u \Concat L_t; \Delta \vdasho
u[t/x]$.

\noindent
We first prove by induction that if a variable $x$ does not occur in
an expression $u$, then $u[t/x] = u$. Then, each SL is proven by
induction on $u$:
\begin{itemize}
\item $(), \New, \Delete, r_n$: both SL by absurdity from the induction hypothesis (IH): no variables in the context.
\item $x\vdasho x$: for both SL, we have $;L_t; \vdasho t$ and $x[t/x] = t$ so we have $;L_t; \vdasho x[t/x]$.
\item $\iota_iv$: for left SL, by IH we have $\Gamma; L_t \Concat L_u; \Delta \vdasho v[t/x]$. Because $(\iota_iv)[t/x] = \iota_i(v[t/x])$, we have $\Gamma; L_t \Concat L_u; \Delta \vdasho (\iota_iv)[t/x]$. The right SL is done similarly, swapping $L_t$ and $L_u$: by IH we have $\Gamma; L_u \Concat L_t; \Delta \vdasho v[t/x]$. Because $(\iota_iv)[t/x] = \iota_i(v[t/x])$, we have $\Gamma; L_u \Concat L_t; \Delta \vdasho (\iota_iv)[t/x]$.
\item $\pi_iv$: similarly, because $(\pi_iv)[t/x] = \pi_i(v[t/x])$ both SL hold by IH.
\item $\langle u_1, u_2 \rangle$: because $\langle u_1, u_2 \rangle[t/x] = \langle u_1[t/x], u_2[t/x] \rangle$, both SL hold by IH.
\item $\lambda y.u_1$: because $\lambda y.(u_1[t/x]) = (\lambda y.u_1)[t/x]$, both SL hold by IH.
\item $(v, w)$: for both SL, given $\Theta \vdasho v$ and $\Theta' \vdasho w$, by case analysis on the three cases in which $\Theta\CtxComp\Theta'$ is well-formed. In the left SL case: \begin{itemize}
    \item $\Theta = \Gamma_v;;$ and $\Theta' = \Gamma_w^1, x; L_u; \Gamma_w^2$: $x$ does not occur in $v$ so $(v, w)[t/x] = (v, w[t/x])$ and by IH on $w$ we have $\Gamma_w^1, x; L_u; \Gamma_w^2 \vdasho w[t/x]$, hence we have $\Gamma_v, \Gamma_w^1; L_t \Concat L_u; \Gamma_w^2 \vdasho (v, w)[t/x]$.
    \item $\Theta = \Gamma_v^1, x; L_u; \Gamma_v^2$ and $\Theta' = \Gamma_w;;$: $(v, w)[t/x] = (v[t/x], w)$ so the SL holds by IH on $\Theta \vdasho v$.
    \item $\Theta = \Gamma_v, x; L_u^1;$ and $\Theta' = ;L_u^2; \Gamma_w$: $(v, w)[t/x] = (v[t/x], w)$ so the SL holds by IH on $v$.
\end{itemize} The right SL is done similarly, swapping $x$ and $L_u$: \begin{itemize}
    \item $\Theta = ;;\Gamma_v$ and $\Theta' = \Gamma_w^1; L_u; x, \Gamma_w^2$: $(v, w)[t/x] = (v, w[t/x])$ so the SL holds by IH on $w$.
    \item $\Theta = \Gamma_v^1; L_u; x, \Gamma_v^2$ and $\Theta' = \Gamma_w$: $(v, w)[t/x] = (v[t/x], w)$ so the SL holds by IH on $v$.
    \item $\Theta = \Gamma_v; L_u^1;$ and $\Theta' = ;L_u^2; x, \Gamma_w$: $(v, w)[t/x] = (v, w[t/x])$ so the SL holds by IH on $w$.
\end{itemize}
\item $\Let y=u_1 \In u_2$: similarly, for both SL we proceed by case analysis on $\Theta \CtxComp \Theta'$. For the left SL: \begin{itemize}
    \item $\Theta = ;;\Gamma_{u2}, y$ and $\Theta' = \Gamma_{u1}^1, x, ;L_u; \Gamma_{u1}^2$: $(\Let y=u_1 \In u_2)[t/x] = (\Let y=u_1 \In u_2[t/x])$ so the SL holds by IH on $u_2$.
    \item $\Theta = \Gamma_{u2}^1, x; L_u; \Gamma_{u2}^2, y$ and $\Theta' = ;;\Gamma_w$: $(\Let y=u_1 \In u_2)[t/x] = (\Let y=u_1[t/x] \In u_2)$ so the SL holds by IH on $u_1$.
    \item $\Theta = \Gamma_{u2}; L_u^1; y$ and $\Theta' = ;L_u^2; x, \Gamma_{u1}$: the concatenation is not well-formed due to $y$ so the SL holds by absurdity.
\end{itemize} The right SL is done similarly by swapping $x$ and $L_u$.
\item $(v, y.u_1, z.u_2)$: For both SL we proceed by case analysis on $\Theta\CtxComp\Theta'\CtxComp\Theta''$, here for left SL: \begin{itemize}
    \item $\Theta = \Gamma_u^1, x; L_u; \Gamma_u^2, \Theta' = ;;\Gamma_v, \Theta'' = ;;\Gamma_u^3$: $(v, y.u_1, z.u_2)[t/x] = (v, y.u_1[t/x], z.u_2[t/x])$ so the SL holds by IH on $\Gamma_u^1, x; L_u; \Gamma_u^2, y, \Gamma_u^3 \vdasho u_1$ and $\Gamma_u^1, x; L_u; \Gamma_u^2, y, \Gamma_u^3 \vdasho u_2$.
    \item $\Theta = \Gamma_u^1, x; L_u^1, \Theta' = ;L_u^2; \Gamma_v, \Theta'' = ;;\Gamma_u^2$: the SL holds by IH on $u_1$ and $u_2$.
    \item $\Theta = ;;\Gamma_u^1, \Theta' = \Gamma_v^1, x; L_u; \Gamma_v^2, \Theta'' = ;;\Gamma_u^2$: the SL holds by IH on $v$.
    \item $\Theta = ;;\Gamma_u^1, \Theta' = \Gamma_v^1, x; L_u^1; \Theta'' = ;L_u^2; \Gamma_u^2$: the SL holds by IH on $v$.
    \item $\Theta = ;;\Gamma_u^1, \Theta' = ;;\Gamma_v, \Theta'' = \Gamma_u^2, x; L_u; \Gamma_u^3$: the SL holds by IH on $u_1$ and $u_2$.
    \item The other cases yield ill-formed concatenated context.
\end{itemize} The right SL is done similarly by swapping $x$ and $L_u$.
\item $\delta (v, (y, z).u_1)$: similarly, by case analysis on $\Theta\CtxComp\Theta'\CtxComp\Theta''$.
\item $\delta (v, ().t)$: by case analysis on $\Theta\CtxComp\Theta'$.
\item $v w$: by case analysis on $\Theta\CtxComp\Theta'$.
\end{itemize}

\mypar{Proof of ordered predicate invariant by reduction (\cref{thm:ord-inv})}
\label{proof:ord-inv}

Show that $\forall c1, \forall c2, \forall L, L \vdasho^C c1 \wedge c1 \rightsquigarrow c2 \Rightarrow L \vdasho^C c2$ by case analysis on reduction steps:
\begin{itemize}
    \item $\langle \Let x^+=t \In u\mid s\mid l \rangle \rightsquigarrow \langle t\mid (x^+.u) \cdotp s\mid l \rangle$: by definition of $L \vdasho^C c1$, we get $;L_t; \vdasho t$, $;L_u; x \vdasho u$ and $L_s \vdasho^S s$ with $L = L_s \Concat L_u \Concat L_t \Concat l$, hence we have $;L_s \Concat L_u; \vdasho (x^+.u) \cdotp s$ and so $L_s \Concat L_u \Concat L_t \Concat l = L \vdasho^C c2$.
    \item $\langle \Let x^-=t \In u\mid s\mid l \rangle \rightsquigarrow \langle u[t/x]\mid s\mid l \rangle$: by definition, we get $;L_t; \vdasho t$, $;L_u; x \vdasho u$ and $L_s \vdasho^S s$ hence by right SL we have $;L_u \Concat L_t; \vdasho u[t/x]$ and so $L \vdasho^C c2$.
    \item $\langle v w\mid s\mid l \rangle \rightsquigarrow \langle v\mid w\cdotp s\mid l \rangle$: by definition, we get $;L_w; \vdasho w$, $;L_v; \vdasho v$ and $L_s \vdasho^S s$, hence we have $L_s \Concat L_w \vdasho^S w\cdotp s$ and so $L \vdasho^C c2$.
    \item $\langle \pi_i t\mid s\mid l \rangle \rightsquigarrow \langle t\mid \pi_i\cdotp s\mid l \rangle$: by definition, we get $;L_t; \vdasho t$ and $L_s \vdasho^S s$, hence we have $L \vdasho^C c2$.
    \item $\langle v\mid (x.t) \cdotp s\mid l \rangle \rightsquigarrow \langle t[v/x]\mid s\mid l \rangle$: by definition, we get $;L_v; \vdasho v$, $;L_t; x \vdasho t$ and $L_s \vdasho^S s$, hence by right SL we have $;L_t \Concat L_v; \vdasho t[v/x]$ and so $L \vdasho^C c2$.
    \item $\langle \lambda x.t\mid v \cdotp s\mid l \rangle \rightsquigarrow \langle t[v/x]\mid s\mid l \rangle$: by definition, we get $L_v \vdasho v$, $x, L_t \vdasho t$ and $L_s \vdasho^S s$, hence by left SL we have $L_v, L_t \vdasho t[v/x]$ and so $L \vdasho^C c2$.
    \item $\langle \langle t_1, t_2 \rangle\mid \pi_i \cdotp s\mid l \rangle \rightsquigarrow \langle t_i\mid s\mid l \rangle$: by definition, we get $L_t \vdasho t_1$, $L_t \vdasho t_2$ and $L_s \vdasho^S s$, hence we have $L \vdasho^C c2$.
    \item $\langle \delta ((v,w), (x,y).t)\mid s\mid l \rangle \rightsquigarrow \langle t[v/x, w/y]\mid s\mid l \rangle$: by definition and case analysis on composition of contexts, we get $;L_v; \vdasho v$, $;L_w; \vdasho w$, $\Theta \vdasho t$ and $L_s \vdasho^S s$ with $\Theta=$ either $;L_t; x, y$ or $x, y; L_t;$. In the first case, by right SL twice we have $;L_t \Concat L_v \Concat L_w; \vdasho t[v/x, w/y]$ and so $L \vdasho^C c2$. Similarly, in the second case by left SL twice we have $;L_v \Concat L_w \Concat L_t; \vdasho t[v/x, w/y]$ and so $L \vdasho^C c2$.
    \item $\langle \delta ((), ().t)\mid s\mid l \rangle \rightsquigarrow \langle t\mid s\mid l \rangle$: by definition, we get $;[]; \vdasho ()$, $;L_t; \vdasho t$ and $L_s \vdasho^S s$, hence we have $L \vdasho^C c2$.
    \item $\langle \delta (\iota_iv, x_1.t_1, x_2.t_2)\mid s\mid l \rangle \rightsquigarrow \langle t_i[x_i/v]\mid s\mid l \rangle$: by definition and case analysis, we get $;L_v; \vdasho v$, $\Theta_i \vdash t_i$ and $L_s \vdasho^S s$ with both $\Theta_i=$ either $;L_t; x_i$ or $x_i; L_t;$. In the first case, by right SL we have $;L_t \Concat L_v; \vdasho t_i[v/x_i]$ and so $L \vdasho^C c2$. Similarly, in the second case by left SL we have $;L_v \Concat L_t; \vdasho t_i[v/x_i]$ and so $L \vdasho^C c2$.
    \item $\langle \New\mid ()\cdotp s\mid r_n\cons l \rangle \rightsquigarrow \langle \iota_1r_n\mid s\mid l \rangle$: by definition, we get $;[]; \vdasho ()$, $;[r_n]; \vdasho r_n$ and $L_s \vdasho^S s$, hence we have $L_s \Concat r_n\cons l = L \vdasho^C c2$.
    \item $\langle \New\mid ()\cdotp s\mid \Nil \rangle \rightsquigarrow \langle \iota_2()\mid s\mid \Nil \rangle$: by definition, we get $L_s = L \vdasho^C c2$.
    \item $\langle \Delete\mid r_n\cdotp s\mid l \rangle \rightsquigarrow \langle ()\mid s\mid r_n\cons l \rangle$: by definition, we get $;[r_n]; \vdasho r_n$, $;[]; \vdasho ()$ and $L_s \vdasho^S s$, hence we have $L_s \Concat r_n\cons l = L \vdasho^C c2$.
\end{itemize}

\mypar{Proof of (\cref{thm:linp}) in the unordered case}
\label{proof:linearity}
We define the indexed predicate $M ; \Gamma \vdashl t$ by induction with $M$ the multiset of resources and $\Gamma$ the set of variables in $t$. Given a linear expression $M ; \Gamma \vdashl t$, we define its multiset of resources $MR(t) \eqdef M$. The concatenation of contexts is always defined: $\Theta=(M_1 ; \Gamma_1) \CtxComp \Theta'=(M_2 ; \Gamma_2) \eqdef M_1, M_2 ; \Gamma_1, \Gamma_2$.

\begin{figure}[tb]
\begin{framed}
\centering
    
\AxiomC{}
\RightLabel{}
\UnaryInfC{$[]; x\vdashl x$}
\DisplayProof
\hfil
\AxiomC{}
\RightLabel{}
\UnaryInfC{$[]; \vdashl ()$}
\DisplayProof
\hfil
\AxiomC{}
\RightLabel{}
\UnaryInfC{$[]; \vdashl \New$}
\DisplayProof
\hfil
\AxiomC{}
\RightLabel{}
\UnaryInfC{$[]; \vdashl \Delete$}
\DisplayProof
\hfil
\AxiomC{}
\RightLabel{}
\UnaryInfC{$[r_n] ; \vdashl r_n$}
\DisplayProof

\vfil
\vRuleSpace

\AxiomC{$\Theta\vdashl v$}
\RightLabel{}
\UnaryInfC{$\Theta\vdashl \iota_iv$}
\DisplayProof
\hfil
\AxiomC{$\Theta\vdashl v$}
\RightLabel{}
\UnaryInfC{$\Theta\vdashl \pi_iv$}
\DisplayProof
\hfil
\AxiomC{$\Theta\vdashl t$}
\AxiomC{$\Theta\vdashl u$}
\RightLabel{}
\BinaryInfC{$\Theta\vdashl \langle t, u \rangle$}
\DisplayProof
\hfil
\AxiomC{$\Theta, x \vdashl t$}
\RightLabel{}
\UnaryInfC{$\Theta \vdashl \lambda x.t$}
\DisplayProof
\hfil
\AxiomC{$\Theta\vdashl v$}
\AxiomC{$\Theta'\vdashl w$}
\RightLabel{}
\BinaryInfC{$\Theta\CtxComp\Theta'\vdashl (v, w)$}
\DisplayProof

\vfil
\vRuleSpace

\AxiomC{$\Theta,x\vdashl u$}
\AxiomC{$\Theta'\vdashl t$}
\RightLabel{}
\BinaryInfC{$\Theta\CtxComp\Theta'\vdashl \Let x = t \In u$}
\DisplayProof
\hfil
\AxiomC{$\Theta,x,y\vdashl t$}
\AxiomC{$\Theta'\vdashl v$}
\RightLabel{}
\BinaryInfC{$\Theta\CtxComp\Theta'\vdashl \delta (v, (x,y).t)$}
\DisplayProof
\hfil
\AxiomC{$\Theta\vdashl t$}
\AxiomC{$\Theta'\vdashl v$}
\RightLabel{}
\BinaryInfC{$\Theta\CtxComp\Theta'\vdashl \delta (v, ().t)$}
\DisplayProof

\vfil
\vRuleSpace

\AxiomC{$\Theta,x\vdashl t$}
\AxiomC{$\Theta,y\vdashl u$}
\AxiomC{$\Theta'\vdashl v$}
\RightLabel{}
\TrinaryInfC{$\Theta\CtxComp\Theta'\vdashl \delta (v, x.t, y.u)$}
\DisplayProof
\hfil
\AxiomC{$\Theta\vdashl w$}
\AxiomC{$\Theta'\vdashl v$}
\RightLabel{}
\BinaryInfC{$\Theta\CtxComp\Theta'\vdashl v w$}
\DisplayProof

\end{framed}
\vspace*{-1.3em}
\caption{Predicate of linear expressions with resources}
\label{linear-rules}
\end{figure}

We can then prove the following substitution lemma:

\begin{lemma}
If $\Theta \vdashl t$ and $\Theta', x \vdashl u$, then $\Theta \CtxComp \Theta' \vdashl u[t/x]$.
\end{lemma}

\begin{proof}
By induction on $u$:
\begin{itemize}
    \item $(), \New, \Delete, r_n$: by absurdity from the induction hypothesis (IH): no variables in the context.
    \item $;x ; \vdashl x$: we have $\Theta \vdashl t$ and $x[t/x] = t$, hence $\Theta \vdashl x[t/x]$.
    \item $\iota_iv$: by IH we get $\Theta \vdashl v[t/x]$ and $(\iota_iv)[t/x] = \iota_i(v[t/x])$, hence $\Theta \vdashl (\iota_iv)[t/x]$.
    \item $\pi_iv$: similarly, because $(\pi_iv)[t/x] = \pi_i(v[t/x])$ the SL holds by IH.
    \item $\langle u_1, u_2 \rangle$: because $\langle u_1, u_2 \rangle[t/x] = \langle u_1[t/x], u_2[t/x] \rangle$, the SL hold by IH.
    \item $\lambda y.u_1$: because $\lambda y.(u_1[t/x]) = (\lambda y.u_1)[t/x]$, both SL hold by IH.
    \item $(v, w)$: by case analysis to split the context $\Theta \CtxComp \Theta'$, whether $x$ is in $v$ or $w$: \begin{itemize}
        \item $\Theta = M_v ; \Gamma_v, x$ and $\Theta' = M_w ; \Gamma_w$: $(v, w)[t/x] = (v[t/x], w)$ so the SL holds by IH on $\Theta \vdashl v$.
        \item $\Theta = M_v ; \Gamma_v$ and $\Theta' = M_w ; \Gamma_w, x$: $(v, w)[t/x] = (v, w[t/x])$ so the SL holds by IH on $\Theta' \vdashl w$.
    \end{itemize}
    \item $\Let y=u_1 \In u_2$: similarly, we proceed by case analysis on the context: \begin{itemize}
        \item $x \in \Theta$: $(\Let y=u_1 \In u_2)[t/x] = (\Let y=u_1 \In u_2[t/x])$ so the SL holds by IH on $\Theta, y \vdashl u_2$.
        \item $x \in \Theta'$: $(\Let y=u_1 \In u_2)[t/x] = (\Let y=u_1[t/x] \In u_2)$ so the SL holds by IH on $\Theta' \vdashl u_1$.
    \end{itemize}
    \item $(v, y.u_1, z.u_2)$: by case analysis on the context: \begin{itemize}
        \item $x \in \Theta$: $(v, y.u_1, z.u_2)[t/x] = (v, y.u_1[t/x], z.u_2[t/x])$ so the SL holds by IH on $\Theta, y \vdashl u_1$ and $\Theta, z \vdashl u_2$.
        \item $x \in \Theta'$: $(v, y.u_1, z.u_2)[t/x] = (v[t/x], y.u_1, z.u_2)$ the SL holds by IH on $\Theta' \vdashl v$.
    \end{itemize}
    \item $\delta (v, (y, z).u_1)$: by case analysis on the context.
    \item $\delta (v, ().t)$: by case analysis on the context.
    \item $v w$: by case analysis on the context.
\end{itemize}
\end{proof}

We then extend $\vdashl$ for stacks and commands, with a multiset of resources and no variables:

\begin{figure}[tb]
\begin{framed}
\centering

\AxiomC{}
\RightLabel{}
\UnaryInfC{$\vdashl^S \star$}
\DisplayProof
\hfil
\AxiomC{$M_v; \vdashl v$}
\AxiomC{$M_s \vdashl^S s$}
\RightLabel{}
\BinaryInfC{$M_s \Concat M_v \vdashl^S v\cdotp s$}
\DisplayProof
\hfil
\AxiomC{$M_t; x \vdashl t$}
\AxiomC{$M_s \vdashl^S s$}
\RightLabel{}
\BinaryInfC{$M_s \Concat M_s \vdashl^S (x^+.t)\cdotp s$}
\DisplayProof
\hfil
\AxiomC{$M_t; \vdashl t$}
\AxiomC{$M_s \vdashl^S s$}
\RightLabel{}
\BinaryInfC{$M_t \Concat M_s \Concat l \vdashl^C \langle t \mid s \mid l \rangle$}
\DisplayProof

\end{framed}
\vspace*{-1.3em}
\caption{Typing rules of ordered stacks and commands}
\end{figure}

\begin{lemma}
Reducing a linear command results in a linear command with the same
multiset of resources, i.e. for all $c_1,c_2,M$ such that $M \vdashl^C
c_1$ and $c_1 \rightsquigarrow c_2$ one has $M \vdashl^C c_2$.
\end{lemma}

\begin{proof}
By induction on reduction steps $c1 \rightsquigarrow c2$:
\begin{itemize}
    \item $\langle \Let x^+=t \In u\mid s\mid l \rangle \rightsquigarrow \langle t\mid (x^+.u) \cdotp s\mid l \rangle$: by definition of $M \vdashl^C c1$, we get $M_t; \vdashl t$, $M_u; x \vdashl u$ and $M_s \vdashl^S s$ with $M = M_s \Concat M_u \Concat M_t \Concat l$, hence we have $M_s \Concat M_u \vdashl (x^+.u) \cdotp s$ and so $M_s \Concat M_u \Concat M_t \Concat l = M \vdashl^C c2$.
    \item $\langle \Let x^-=t \In u\mid s\mid l \rangle \rightsquigarrow \langle u[t/x]\mid s\mid l \rangle$: by definition, we get $M_t; \vdashl t$, $M_u; x \vdashl u$ and $M_s \vdashl^S s$, hence by SL we have $M_t \Concat M_u ; \vdashl t[u/x]$ and so $M \vdashl^C c2$.
    \item $\langle \pi_i t\mid s\mid l \rangle \rightsquigarrow \langle t\mid \pi_i\cdotp s\mid l \rangle$: by definition, we get $M_t; \vdashl t$ and $M_s \vdashl^S s$, hence $M \vdashl^C c2$.
    \item $\langle v\mid (x.t) \cdotp s\mid l \rangle \rightsquigarrow \langle t[v/x]\mid s\mid l \rangle$: by definition, we get $M_v; \vdashl v$, $M_t; x \vdashl t$ and $M_s \vdashl^S s$, hence by SL we have $M_t \Concat M_v; \vdashl t[v/x]$ and so $M \vdashl^C c2$.
    \item $\langle \lambda x.t\mid v \cdotp s\mid l \rangle \rightsquigarrow \langle t[v/x]\mid s\mid l \rangle$: by definition, we get $M_t; x \vdashl t$, $M_v; \vdashl v$ and $M_s \vdashl^S s$, hence by SL we have $M_t \Concat M_v; \vdashl t[v/x]$ and so $M \vdashl^C c2$.
    \item $\langle \langle t_1, t_2 \rangle\mid \pi_i \cdotp s\mid l \rangle \rightsquigarrow \langle t_i\mid s\mid l \rangle$: by definition, we get $M_t; \vdashl t_1$, $M_t; \vdashl t_2$ and $M_s \vdashl^S s$, hence $M \vdashl^C c2$.
    \item $\langle \delta ((v,w), (x,y).t)\mid s\mid l \rangle \rightsquigarrow \langle t[v/x, w/y]\mid s\mid l \rangle$: by definition, we get $M_v; \vdashl v$, $M_w; \vdashl w$, $M_t; x, y \vdashl t$ and $M_s \vdashl^S s$, hence by SL twice we have $M_t \Concat M_v \Concat M_w; \vdashl t[v/x][w/y] = t[v/x, w/y]$ and so $M \vdashl^C c2$.
    \item $\langle \delta ((), ().t)\mid s\mid l \rangle \rightsquigarrow \langle t\mid s\mid l \rangle$: by definition, we get $M_t; \vdashl t$ and $M_s \vdashl^S s$, hence $M \vdashl^C c2$.
    \item $\langle \delta (\iota_iv, x_1.t_1, x_2.t_2)\mid s\mid l \rangle \rightsquigarrow \langle t_i[x_i/v]\mid s\mid l \rangle$: by definition, we get $M_v; \vdashl v$, $M_t; x_1 \vdashl t_1$, $M_t; x_2 \vdashl t_2$ and $M_s \vdashl^S s$, hence by SL we have $M_t \Concat M_v; \vdashl t_i[v/x_i]$ and so $M \vdashl^C c2$.
    \item $\langle \New\mid ()\cdotp s\mid r_n\cons l \rangle \rightsquigarrow \langle \iota_1r_n\mid s\mid l \rangle$: by definition, we get $; \vdashl \New$, $; \vdashl ()$ and $M_s \vdashl^S s$, hence $M_s \Concat r_n\cons l = M \vdashl^C c2$.
    \item $\langle \New\mid ()\cdotp s\mid \Nil \rangle \rightsquigarrow \langle \iota_2()\mid s\mid \Nil \rangle$: by definition, we get $; \vdashl \New$, $; \vdashl ()$ and $M_s \vdashl^S s$, hence $M_s = M \vdashl^C c2$.
    \item $\langle \Delete\mid r_n\cdotp s\mid l \rangle \rightsquigarrow \langle ()\mid s\mid r_n\cons l \rangle$: by definition, we get $; \vdashl \Delete$, $r_n ; \vdashl r_n$ and $M_s \vdashl^S s$, hence $M_s \Concat r_n\cons l = M \vdashl^C c2$.
\end{itemize}
\end{proof}

Finally, we show that $\forall \vdash t: P \in Pr, \forall v, \forall l, \forall l', \langle t \mid \star \mid l \rangle \rightsquigarrow \langle v \mid \star \mid l' \rangle \Rightarrow \exists \sigma, l' = \sigma(l)$.

\begin{proof}
By progress and subject reduction, we have that $\vdash v: W$. By
induction on the typing derivation, we have that $\vdash t:
A \Rightarrow ; \vdashl t$, hence $l \vdashl^C \langle
t \mid \star \mid l \rangle$. By the previous lemma, we get
$l \vdashl^C \langle v \mid \star \mid l' \rangle$. Because $W$ is
central, $v$ is a final value, so it does not contain any resource.
Hence, $l$ and $l'$ are obtained from the same multiset, so
$\exists \sigma, l' = \sigma(l)$.
\end{proof}

\section{Details: translation of the resource CBPV}

\mypar{Proof of well-typed translation}
\label{proof:translation}

All terms must be translated to values. Recall that $\Down A \eqdef A \With 1$, hence the type of translated expressions and negative values is negative, so they have no value restriction. So, we must only check the value restriction when translating positive values.

Contexts are translated by translating each type separately, so $(\Gamma, \Delta)^+ = \Gamma^+, \Delta^+$.

Values of central types are translated to the same value and type in \LinLang, so $\llbracket \vdash \NewFail:E \rrbracket = {\vdash \NewFail:E}$. Moreover, because \AffLang is an extension of \LinLang, it can use linear swaps for exceptions.

We proceed by cases:
\begin{itemize}
\item
$\llbracket x \vdash x:A \rrbracket$ must be a value of type $A^+ \vdash A^+$.
Hence, $x$ is well-typed.
\item
$\llbracket \vdash \DropB: \Lmap{A}{1} \rrbracket$ must be a value of type $\vdash (\Lmap{A^+}{1}) \With 1$.
Hence, $\langle \Drop_A, () \rangle$ is well-typed.
\item
$\llbracket \Gamma \vdash \Coerc(v): P \rrbracket$ must be a value of type $\Gamma^+  \vdash (P^+ \oplus E) \With I$.
By IH, we have $\Gamma^+  \vdash \llbracket v \rrbracket : P^+$.
Hence, $\langle \iota_1 \llbracket v \rrbracket , \DropCtx_\Gamma \rangle$ is well-typed.
\item
$\llbracket \Gamma,x:A,\Delta \vdash \Move(x) \In t:C \rrbracket$ must be a value of type $\Gamma^+, \Delta^+, x:A^+ \vdash \Down C^-$.
By IH, we have $\Gamma^+ , x:A^+, y:B^+, \Delta^+ \vdash \llbracket t \rrbracket :\Down C^-$.
It's the only rule where we allow ourselves to use \LinLang structural rule to swap premises.
Hence, $\llbracket t \rrbracket $ is well-typed.
\item
$\llbracket \vdash \New : \Lmap{1}{R} \rrbracket$ must be a value of type $\vdash \Lmap{1}{R \oplus E}$. We have $\vdash \NewFail: E$, so $\Let x = \New () \In \delta(x, r. \iota_1 r, i.i; \iota_2 \NewFail)$ is well-typed.
\item
$\llbracket \Gamma, \Delta \vdash \Let x=(v:A) \In t:B \rrbracket$ must be a value of type $\Gamma^+, \Delta^+ \vdash \Down B^-$.
By IH, we have $\Delta^+ \vdash \llbracket v \rrbracket : A^+$ and $\Gamma^+ , x:A^+ \vdash \llbracket t \rrbracket : \Down B^-$.
Hence, $\Let x = \llbracket v \rrbracket  \In \llbracket t \rrbracket$  is well-typed.
\item
$\llbracket \Gamma, \Delta \vdash \Let x=(t:P) \In u:A \rrbracket$ must be a value of type $\Gamma^+, \Delta^+ \vdash A^- \With I$.
By IH, we have $\Delta^+ \vdash \llbracket t \rrbracket : (P^+ \oplus E) \With I$ and $\Gamma^+ , x:P^+ \vdash \llbracket u \rrbracket : A^- \With I$. Then, $\Raise^{A \With I}_\Gamma(e)$ is of type $\Gamma^+ \vdash A^- \With I$, so $\Let s = \pi_1\llbracket t \rrbracket  \In \delta(s, x. \llbracket u \rrbracket , e. \Raise^{A \With I}_\Gamma(e))$ is well-typed.
\item
$\llbracket \vdash \RaiseB : \Lmap{E}{A} \rrbracket$ must be a value of type $\vdash (\Lmap{E}{A^-}) \With I$.
Hence, $\langle \lambda e. \Raise^A_{\star}(e), () \rangle$ is well-typed.
\item
$\llbracket \Gamma, \Delta \vdash \Try x \Leftarrow (t:P) \In u \Unless e \Rightarrow u' : B \rrbracket$ must be a value of type $\Gamma^+, \Delta^+\vdash \Down B^-$.
By IH, we have:
\begin{itemize}
  \item $\Delta^+ \vdash \llbracket t \rrbracket : (P^+ \oplus E) \With I$.
  \item $\Gamma^+ , x:P^+ \vdash \llbracket u \rrbracket : \Down B^-$.
  \item $\Gamma^+ , e:E \vdash \llbracket u' \rrbracket : \Down B^-$.
\end{itemize}
Hence, $\Let s = \pi_1\llbracket t \rrbracket  \In \delta(s, x. \llbracket u \rrbracket , e. \llbracket u' \rrbracket )$ is well-typed.
\item
$\llbracket \Gamma, \Delta \vdash (v, w): A\otimes B \rrbracket$ must be a value of type $\Gamma^+, \Delta^+ \vdash A^+\otimes B^+$.
By IH, we have $\Gamma^+  \vdash \llbracket v \rrbracket : A^+$ and $\Delta^+ \vdash \llbracket w \rrbracket : B^+$.
Hence, $(\llbracket v \rrbracket , \llbracket w \rrbracket )$ is well-typed.
\item
$\llbracket \Gamma, \Delta, \Gamma' \vdash \delta(v, (x, y).t): C \rrbracket$ must be a value of type $\Gamma^+, \Delta^+, \Gamma'^+  \vdash \Down C^-$.
By IH, we have $\Delta^+ \vdash \llbracket v \rrbracket : A^+\otimes B^+$ and $\Gamma^+ , x:A^+, y:B^+, \Gamma'^+ \vdash \llbracket t \rrbracket : \Down C^-$.
Hence, $\delta(\llbracket v \rrbracket , (x, y). \llbracket t \rrbracket )$ is well-typed.
\item
$\llbracket \vdash (): 1 \rrbracket$ must be a value of type $\vdash 1$.
Hence, $()$ is well-typed.
\item
$\llbracket \Gamma, \Delta, \Gamma' \vdash \delta(v, ().t): C \rrbracket$ must be a value of type $\Gamma^+, \Delta^+, \Gamma'^+  \vdash \Down C^-$.
By IH, we have $\Delta^+ \vdash \llbracket v \rrbracket : 1$ and $\Gamma^+ , \Gamma'^+ \vdash \llbracket t \rrbracket : \Down C^-$.
Hence, $\delta(\llbracket v \rrbracket , (). \llbracket t \rrbracket )$ is well-typed.
\item
$\llbracket \Gamma \vdash \iota_1v: A\oplus B \rrbracket$ must be a value of type $\Gamma^+  \vdash A^+ \oplus B^+$.
By IH, we have $\Gamma^+  \vdash \llbracket v \rrbracket : A^+$.
Hence, $\iota_1\llbracket v \rrbracket$ is well-typed.
\item
$\llbracket \Gamma \vdash \iota_2v: A\oplus B \rrbracket$ must be a value of type $\Gamma^+  \vdash A^+ \oplus B^+$.
By IH, we have $\Gamma^+  \vdash \llbracket v \rrbracket : B^+$.
Hence, $\iota_2\llbracket v \rrbracket$ is well-typed.
\item
$\llbracket \Gamma, \Delta, \Gamma' \vdash \delta(v, x.t, y.u): C \rrbracket$ must be a value of type $\Gamma^+, \Delta^+, \Gamma'^+  \vdash \Down C^-$.
By IH, we have:
\begin{itemize}
  \item $\Delta^+ \vdash \llbracket v \rrbracket : A^+ \oplus B^+$.
  \item $\Gamma^+ , x:A^+, \Gamma'^+ \vdash \llbracket t \rrbracket : \Down C^-$.
  \item $\Gamma^+ , y:B^+, \Gamma'^+ \vdash \llbracket u \rrbracket : \Down C^-$.
\end{itemize}
Hence, $\delta(\llbracket v \rrbracket , x.\llbracket t \rrbracket , y.\llbracket u \rrbracket )$ is well-typed.
\item
$\llbracket \Gamma \vdash \lambda x.t: \Lmap{A}{B} \rrbracket$ must be a value of type $\Gamma^+  \vdash (\Lmap{A^+}{B^-}) \With I$.
By IH, we have $x:A^+, \Gamma^+  \vdash \llbracket t \rrbracket : B^- \With I$.
Hence, $\langle \lambda x. \pi_1\llbracket t \rrbracket , \DropCtx_\Gamma \rangle$ is well-typed.
\item
$\llbracket \Gamma, \Delta \vdash v w: B \rrbracket$ must be a value of type $\Gamma^+, \Delta^+ \vdash \Down B^-$.
By IH, we have $\Gamma^+  \vdash \llbracket w \rrbracket : A^+$ and $\Gamma^+  \vdash \llbracket v \rrbracket : (\Lmap{A^+}{B^-}) \With I$.
Hence, $\langle (\pi_1\llbracket v \rrbracket ) \llbracket w \rrbracket, \DropCtx_{\Gamma, \Delta} \rangle$ is well-typed.
\item
$\llbracket \Gamma \vdash \langle t, u \rangle: A \With B \rrbracket$ must be a value of type $\Gamma^+  \vdash (A^- \With B^-) \With I$.
By IH, we have $\Gamma^+  \vdash \llbracket t \rrbracket : A^- \With I$ and $\Gamma^+  \vdash \llbracket u \rrbracket : B^- \With I$.
Hence, $\langle \langle \pi_1\llbracket t \rrbracket , \pi_1\llbracket u \rrbracket  \rangle, \DropCtx_\Gamma \rangle$ is well-typed.
\item
$\llbracket \Gamma \vdash \pi_1v: A \rrbracket$ must be a value of type $\Gamma^+  \vdash A^- \With I$.
By IH, we have $\Gamma^+  \vdash \llbracket v \rrbracket : (A^- \With B^-) \With I$.
Hence, $\langle \pi_1\pi_1\llbracket v \rrbracket , \DropCtx_\Gamma \rrbracket  \rangle$ is well-typed.
\item
$\llbracket \Gamma \vdash \pi_2v: B \rrbracket$ must be a value of type $\Gamma^+  \vdash B^- \With I$.
By IH, we have $\Gamma^+  \vdash \llbracket v \rrbracket : (A^- \With B^-) \With I$.
Hence, $\langle \pi_2\pi_1\llbracket v \rrbracket , \DropCtx_\Gamma \rrbracket  \rangle$ is well-typed.
\end{itemize}

\section{Details: presheaf model}

\label{sec:cat-interp}
\mypar{Interpretation of types and terms}

Resources appearing in terms of the operational semantics are
interpreted as variables of type $R$.

Each type has a positive and negative interpretation:
\begin{itemize}
    \item $1^+ \eqdef I, (A \otimes B)^+ \eqdef A^+ \otimes B^+, (A \oplus B)^+ \eqdef A^+ \oplus B^+, N^+ \eqdef G(N^-)$
    \item $(\Lmap{A}{B})^- \eqdef \Lmap{A^+}{B^-}, (A \With B)^- \eqdef A^- \times B^-, P^- \eqdef F(P^+)$
\end{itemize}

A context with resources $\Gamma; [r_1 \dots r_n]; \Delta$ is interpreted positively by $\Gamma^+ \otimes R^n \otimes \Delta^+$ with $R^{n+1} \eqdef R^n \otimes R, R^0 \eqdef I, (x: A, \Gamma)^+ \eqdef A^+ \otimes \Gamma^+$. Lists of resources are then written $L_x$ and their translation $L_x^+$.

Each judgement kind with resources have an interpretation:
\begin{itemize}
    \item Values $\Theta \vdasho v: A$ in $\Ccat(\Theta^+, A^+)$.
    \item Expressions $\Theta \vdasho t: A$ in $\Ccat(\Theta^+, GA^-)$.
    \item Stacks $Ls; s: A \vdasho B$ in $\Ccat(Ls^+ \otimes A^-, B^-)$.
    \item Commands $Lc \vdasho c: A$ in $\Ccat(Lc^+, A^-)$.
\end{itemize}

To interpret terms, types in the derivation of $\vdasho$ are made explicit. In particular, stacks $L \vdasho^S s$ are written $L; s: A \vdasho^S B$.

Composition is written in the diagrammatic order with $;$. $\eta$ is the unit of $GF$. $f^*$ is the adjoint of $f$ under $F \dashv G$. $f^{str_\Gamma}$ precomposes $f$ with the isomorphism $\Gamma \otimes FA \cong F(\Gamma \otimes A)$ in the relevant direction. $f^{\lambda^{\LmapSym/\RmapSym}}$ is the adjoint of $f$ under $\otimes \dashv \LmapSym$ (respectively $\RmapSym$), and $ev^{\LmapSym/\RmapSym}$ the evaluation morphism of this adjunction. $dis$ is the canonical morphism $\Gamma \otimes (A \oplus B) \otimes \Gamma' \to (\Gamma \otimes A \otimes \Gamma') \oplus (\Gamma \otimes B \otimes \Gamma')$ obtained from the fact that the category is monoidal distributive, because it is biclosed. Apart from $str$, associators and unitors of the monoidal product are left implicit.

Values have explicit coercions inserted on positive values where expressions are expected, written $\Coerc(v)$ below.
\begin{itemize}
    \item $\llbracket \Theta \vdasho \Coerc(v):P \rrbracket \eqdef \llbracket v \rrbracket; \eta: \Ccat(\Theta^+, GFP^+)$
    \item $\llbracket x:A \vdasho x:A \rrbracket \eqdef id_{A^+}: \Ccat(A^+, A^+)$
    \item $\llbracket r:R \vdasho r:R \rrbracket \eqdef id_R: \Ccat(R, R)$
    \item $\llbracket \vdasho ():1 \rrbracket \eqdef id_I: \Ccat(I, I)$
    \item $\llbracket \Theta, \Theta' \vdasho (v, w): A \otimes B \rrbracket \eqdef \llbracket v \rrbracket \otimes \llbracket w \rrbracket: \Ccat(\Theta^+ \otimes \Theta'^+, A^+ \otimes B^+)$
    \item $\llbracket \Theta \vdasho \iota_1v: A \oplus B \rrbracket \eqdef \llbracket v \rrbracket;inl: \Ccat(\Theta^+, A^+ \oplus B^+)$
    \item $\llbracket \Theta \vdasho \iota_2v: A \oplus B \rrbracket \eqdef \llbracket v \rrbracket;inr: \Ccat(\Theta^+, A^+ \oplus B^+)$
    \item $\llbracket \Theta, \Theta', \Theta'' \vdasho \delta(v, ().t): C \rrbracket \eqdef \llbracket t \rrbracket: \Ccat(\Theta^+ \otimes \Theta'^+ \otimes \Theta^+, GC^-)$
    \item $\llbracket \Theta, \Theta', \Theta'' \vdasho \delta(v, (x,y).t): C \rrbracket \eqdef (id_{\Theta^+} \otimes \llbracket v \rrbracket \otimes id_{\Theta''^+});\llbracket t \rrbracket: \Ccat(\Theta^+ \otimes \Theta'^+ \otimes \Theta^+, GC^-)$
    \item $\llbracket \Theta, \Theta', \Theta'' \vdasho \delta(v, x.t, y.u): C \rrbracket \eqdef (id_{\Theta^+} \otimes \llbracket v \rrbracket \otimes id_{\Theta''^+});dis;(\llbracket t \rrbracket \mid \llbracket u \rrbracket): \Ccat(\Theta^+ \otimes \Theta'^+ \otimes \Theta^+, GC^-)$
    \item $\llbracket \Theta \vdasho \lambda x.t: \Lmap{A}{B} \rrbracket \eqdef \llbracket t \rrbracket^{*;str_{\Theta^+};\lambda^{\LmapSym};*}: \Ccat(\Theta^+, G(\Lmap{A^+}{B^-}))$
    \item $\llbracket \Theta \vdasho \langle t, u \rangle \rrbracket \eqdef (\llbracket t \rrbracket^* \times \llbracket u \rrbracket^*)^*: \Ccat(\Theta^+, G(A^- \With B^-))$
    \item $\llbracket \Theta, \Theta' \vdasho v w: B \rrbracket \eqdef ((\llbracket w \rrbracket \otimes \llbracket v \rrbracket^*);ev)^{str_{\Theta^+};*}: \Ccat(\Theta^+ \otimes \Theta'^+, GB^-)$
    \item $\llbracket \Theta \vdasho \pi_1 v: A \rrbracket \eqdef (v^*;\pi_1)^*: \Ccat(\Theta^+, GA^-)$
    \item $\llbracket \Theta \vdasho \pi_2 v: B \rrbracket \eqdef (v^*;\pi_2)^*: \Ccat(\Theta^+, GB^-)$
    \item $\llbracket \Theta, \Theta' \vdasho \Let x=(t:P) \In u \rrbracket \eqdef ((id_{\Theta^+} \otimes \llbracket t \rrbracket^*);\llbracket u \rrbracket^{*;str_{\Theta^+}})^{str_{\Theta^+};*}: \Ccat(\Theta^+ \otimes \Theta'^+, GB^-)$
    \item $\llbracket \Theta, \Theta' \vdasho \Let x=(t:N) \In u \rrbracket \eqdef (id_{\Theta^+} \otimes \llbracket t \rrbracket);\llbracket u \rrbracket: \Ccat(\Theta^+ \otimes \Theta'^+, GB^-)$
    \item $\llbracket \vdasho \New: \Lmap{1}{R \oplus 1} \rrbracket \eqdef l \mapsto i \mapsto \Match l \{ x\cons t \mapsto (inr(x), t) \mid [] \mapsto (inr(()), []) \} : \Ccat(I, G(\Lmap{I}{F(R \oplus I)}))$
\end{itemize}

Stacks are interpreted as such:
\begin{itemize}
    \item $\llbracket [];\star:A \vdasho A \rrbracket \eqdef id_{A^-}: \Ccat(A^-, A^-)$
    \item $\llbracket Ls; \pi_1 \cdotp s: A\With B \vdasho C \rrbracket \eqdef \pi_1; \llbracket s \rrbracket: \Ccat(Ls^+ \otimes (A^- \With B^-), C^-)$
    \item $\llbracket Ls; \pi_2 \cdotp s: A\With B \vdasho C \rrbracket \eqdef \pi_2; \llbracket s \rrbracket: \Ccat(Ls^+ \otimes (A^- \With B^-), C^-)$
    \item $\llbracket Ls \Concat Lt; v \cdotp s: \Lmap{A}{B} \vdasho C \rrbracket \eqdef (id_{Ls^+} \otimes ((\llbracket v \rrbracket \otimes id_{\Lmap{A^+}{B^-}}); ev^{\LmapSym})); \llbracket s \rrbracket: \Ccat(Ls^+ \otimes Lt^+ \otimes (\Lmap{A^+}{B^-}), C^-)$
    \item $\llbracket Ls \Concat Lt; (x^+.t) \cdotp s: A \vdasho C \rrbracket \eqdef (id_{Ls^+} \otimes \llbracket t \rrbracket^{*;str_{Lt^+}}); \llbracket s \rrbracket: \Ccat(Ls^+ \otimes Lt^+ \otimes FA^+, C^-)$
\end{itemize}

Finally, $\llbracket Ls \Concat Lt \Concat Ll \vdasho \langle t \mid s \mid l \rangle \rrbracket \eqdef (((id_{Ls^+} \otimes \llbracket t \rrbracket^*); \llbracket s \rrbracket)^{str_{Ls^+};*} \otimes (!: Ll \to [R])); ev^{\RmapSym}$.

\mypar{Proof of (\cref{lem:cat-subst})}
\label{proof:cat-subst}

\begin{proof}
By induction on the derivation of $t$ with explicit coercions. Uses naturality of $\otimes$ associativity left implicit, as well as other transformations ($*, str, \times, \lambda$) that interpret rules of negative connectives. For $*$, it means in particular that $f;(g^*; h)^* = (Ff;(g^*;h))^* = ((f;g)^*;h)^*$:
\begin{itemize}
    \item $\llbracket \Coerc(w)[v/x] \rrbracket = \llbracket w[v/x] \rrbracket;\eta = \llbracket \Coerc(w[v/x])$ by induction hypothesis.
    \item $\llbracket x[v/x] \rrbracket = \llbracket v \rrbracket$.
    \item No variables in $r$, $\New$ nor $()$.
    \item $\llbracket \Gamma, \Delta \vdasho (v, w) \rrbracket$: either $x \in \Gamma$ or $x \in \Delta$, in which cases we need to show $\llbracket (w, w')[v/x] \rrbracket = \llbracket (w[v/x], w') \rrbracket$ or $\llbracket (w[v/x], w') \rrbracket$. Those hold by IH.
    \item $\llbracket (\iota_iw)[v/x] \rrbracket = \llbracket \iota_i(w[v/x]) \rrbracket$ by IH.
    \item $\llbracket \delta(w, ().t)[v/x] \rrbracket$: split case on $v \in \Gamma, v \in \Delta$ or $v \in \Gamma'$. In each case, the SL holds by IH.
    \item $\llbracket \delta(w, (y, z).t)[v/x] \rrbracket$: same split case. In each case, the SL holds by IH.
    \item $\llbracket \delta(w, y.t, z.u)[v/x] \rrbracket$: same split case. In each case, the SL holds by IH.
    \item $\langle t, u \rangle$: for $\langle t, u \rangle[v/x] = \langle t[v/x], u[v/x] \rangle$, we need to show that $(id_{\Gamma^+} \otimes \llbracket v \rrbracket \otimes id_{\Gamma'^+}); (\llbracket t \rrbracket^* \times \llbracket u \rrbracket^*)^*$ = $(((id_{\Gamma^+} \otimes \llbracket v \rrbracket \otimes id_{\Gamma'^+});\llbracket t \rrbracket)^* \times ((id_{\Gamma^+} \otimes \llbracket v \rrbracket \otimes id_{\Gamma'^+});\llbracket u \rrbracket)^*)^*$. Hence, the SL holds by IH and naturality of $*$ and $\times$.
    \item $\pi_iv$: similarly, because $(\pi_iv)[t/x] = \pi_i(v[t/x])$ the SL holds by IH and naturality of $*$.
    \item $\lambda y.u_1$: because $\lambda y.(u_1[t/x]) = (\lambda y.u_1)[t/x]$, both SL hold by IH and naturality of $*$ and $str$.
    \item $w w'$: split case on whether we have $w[v/x]=w$ or $w'[v/x]=w'$, in both cases it holds by IH and naturality of $*, str$ and $\lambda$.
    \item $\llbracket \Gamma, \Delta \vdasho (\Let y=(t:P) \In u)[v/x] \rrbracket$: split case on $v \in \Gamma$ or $v \in \Delta$, in both cases it holds by SL and naturality of $*$ and $str$.
    \item $\llbracket \Gamma, \Delta \vdasho (\Let y=(t:N) \In u)[v/x] \rrbracket$: same split case. In both cases it holds by SL.
\end{itemize}
\end{proof}

\mypar{Proof of (\cref{thm:cat-sound})}
\label{proof:cat-sound}

\begin{proof}
By case analysis on reduction steps: when only $t$ changes, we show that expressions from both commands are equal. Otherwise, when the stack also changes, we show that $(Ls^+ \otimes \llbracket t \rrbracket^*); \llbracket s \rrbracket$ from both command are equal. Finally, we consider the whole command interpretation pontwise for $\New$ and $\Delete$ reduction rules.
\begin{itemize}
    \item $\langle \Let x^-=v \In u\mid s\mid l \rangle \rightsquigarrow \langle u[t/x]\mid s\mid l \rangle$: we need to show that $\llbracket \Let x^-=v \In u \rrbracket = \llbracket u[v/x] \rrbracket$. Both expressions are equal by SL and definition of binding a negative value, that is precomposition.
    \item $\langle \Let x^+=t \In u\mid s\mid l \rangle \rightsquigarrow \langle t\mid (x^+.u) \cdotp s\mid l \rangle$: we need to show that $(id_{Ls^+} \otimes ((id_{Lu^+} \otimes \llbracket t \rrbracket^*);\llbracket u \rrbracket^{*;str_{Lu^+}})^{str_{Lu^+};*;*}); \llbracket s \rrbracket = (id_{Ls^+} \otimes id_{Lu^+} \otimes \llbracket t \rrbracket^*); (id_{Ls^+} \otimes \llbracket u \rrbracket^{*;str_{Lu^+}}); \llbracket s \rrbracket$. It holds after cancelling the two $*$, re-associating parentheses with the first $str$ of the left expression then distributing $\otimes$.
    \item $\langle v w\mid s\mid l \rangle \rightsquigarrow \langle v\mid w\cdotp s\mid l \rangle$: holds after cancelling $*;*$.
    \item $\langle \pi_i t\mid s\mid l \rangle \rightsquigarrow \langle t\mid \pi_i\cdotp s\mid l \rangle$: holds after cancelling $*;*$ on the left expression.
    \item $\langle v\mid (x.t) \cdotp s\mid l \rangle \rightsquigarrow \langle t[v/x]\mid s\mid l \rangle$: $\llbracket \Coerc(v) \rrbracket^* = F\llbracket v \rrbracket$, then it holds by SL.
    \item $\langle \lambda x.t\mid v \cdotp s\mid l \rangle \rightsquigarrow \langle t[v/x]\mid s\mid l \rangle$: holds after cancelling $*;*$ and computing $\lambda$ with $ev$ on the left expression, then applying SL.
    \item $\langle \langle t_1, t_2 \rangle\mid \pi_i \cdotp s\mid l \rangle \rightsquigarrow \langle t_i\mid s\mid l \rangle$: holds after cancelling $*;*$ and computing $\pi_i$ on the left expression.
    \item $\langle \delta ((v,w), (x,y).t)\mid s\mid l \rangle \rightsquigarrow \langle t[v/x, w/y]\mid s\mid l \rangle$: both command expressions are equal by applying SL twice.
    \item $\langle \delta ((), ().t)\mid s\mid l \rangle \rightsquigarrow \langle t\mid s\mid l \rangle$: both command expressions are equal by unit laws.
    \item $\langle \delta (\iota_iv, x_1.t_1, x_2.t_2)\mid s\mid l \rangle \rightsquigarrow \langle t_i[x_i/v]\mid s\mid l \rangle$: the left expression with SL is equal to the right after computing the distribution morphism and $inl/inr$ (for each $i$).
    \item $\langle \New\mid ()\cdotp s\mid r_n\cons l \rangle \rightsquigarrow \langle \iota_1r_n\mid s\mid l \rangle$: by definition of popping elements from a non-empty list of resources.
    \item $\langle \New\mid ()\cdotp s\mid \Nil \rangle \rightsquigarrow \langle \iota_2()\mid s\mid \Nil \rangle$: by definition of popping elements from an empty list of resources.
    \item $\langle \Delete\mid r_n\cdotp s\mid l \rangle \rightsquigarrow \langle ()\mid s\mid r_n\cons l \rangle$: by definition of pushing an element in a list of resources.
\end{itemize}
\end{proof}

\mypar{Proof that centre coincides with resource-free objects, and objects with trivial destructors}
\label{proof:movable}

We will show the bi-implications $1 \iff 2$ and $2 \iff 3$ for the following properties on object $E \in \Ccat$ (see \cref{thm:movable}):
\begin{enumerate}
  \item[1] $E$ is in the Drinfeld centre of $\Ccat$.
  \item[2] For all $l$ such that $l \ne []$, $E(l) = \emptyset$.
  \item[3] There is a morphism $\Ccat(E, I)$.
\end{enumerate}

Recall that a morphism $\Ccat(A, B)$ is a function family $\prod_l A(l) \to B(l)$.

$(2 \implies 3)$ Resource-free objects have a trivial destructor:

We define the family of functions $d: \prod_l E(l) \to I(l)$ by case on $l$:
\begin{itemize}
  \item If $l=[]$, then $I([]) = \{\star\}$ so we pick the unique function to it.
  \item Otherwise, $E(l) = \emptyset$ so we pick the unique function from it.
\end{itemize}

$(3 \implies 2)$ Objects with a trivial destructor have no resources:

There is a function family $\forall l, E(l) \to I(l)$.
If $l=[], I(l)=\emptyset$ so it must be the case that $E(l)=\emptyset$.

$(1 \implies 2)$ Objects in the Drinfeld centre have no resources.
Such objects are pairs $E: \Ccat, \theta: E \otimes- \cong - \otimes
E)$ such that
\[\theta_{A \otimes B} = (\theta_A \otimes
id);(id \otimes \theta_B)\;.
\]
So, we have a bijection family
\[
\prod_A \prod_l \prod_{j \Concat k=l}, \sum_{j' \Concat k'= l} E(j) \times A(k) \cong A(j') \times E(k')\;.
\]

Suppose that there exists an inhabited list $r \cons t$ such that $E(r \cons t)$ is inhabited. We will show that no such bijection family exists.

We fix a resource $r' \ne r$.
We instantiate $\theta$ with $A(t) \eqdef \{\star\}, A(-) \eqdef \emptyset, j \eqdef r \cons t, k \eqdef [r']$. We must show that for all $j', k'$ such that $j' \Concat k' = r \cons t \Concat [r']$, there is no function $E(r \cons t) \times A([r']) \to A(j') \times E(k')$ : because $E(r \cons t)$ and $A([r'])$ are inhabited, $A(j') \times E(k')$ needs to be inhabited. However, $j'$ first element must be $r$, so $j'\ne [r']$. Hence $A(j') = \emptyset$, so the product is empty.

$(2 \implies 1)$ We show that resource-free objects are in the
Drinfeld centre.
Given $E$ resource-free, we define $\theta: \prod_A \prod_l \prod_{j \Concat k = l} \sum_{j' \Concat k'= l} E(j) \times A(k) \cong A(j') \times E(k')$ by case on $j$:
\begin{itemize}
  \item If $j=[]$ then $k=l$, so we pick $j'=k, k'=[]$ and the function \[\Swap_{E([]), A(k)}: E([]) \times A(k) \cong A(k) \times E([])\].
  \item Otherwise $E(j)=\emptyset$, so we pick $j'\eqdef j, k'\eqdef k$ and the unique function from $E(j) \times A(k)$.
\end{itemize}
We are left to check naturality and the centre additional condition. If $j\ne[]$, they hold by universal property, so we are left with the first case. $\Swap$ is natural in both arguments, so it only remains to show that \[\theta_{A \otimes B} = (\theta_A \otimes id);(id \otimes \theta_B)\;.\]

Those morphisms are function families \[\prod_{l} \prod_{i \Concat j \Concat k = l} \sum_{i' \Concat j' \Concat k' = l} E(i) \times A(j) \times B(k) \cong A(i') \times B(j') \times E(k')\]. We need to check that they are pointwise equal.
Given $f: A \to B$ and $g: C \to D$, the Day convolution morphism
\[
f \otimes g: \prod_{l} \prod_{j \Concat k = l} \sum_{j' \Concat
k'= l} A(j) \times C(k) \to B(j') \times D(k')\] equals $f(j) \times
g(k)$ by picking $j'\eqdef j, k'\eqdef k$.
By definition, we thus have
\begin{align*}
&(\theta_A \otimes id);(id \otimes \theta_B)(l = i \Concat j \Concat k)\\
&= (\Swap_{E(i), A(j)} \times id_{B(k)});(id_{A(j)} \times \Swap_{E(i), B(k)})\\
&= \Swap_{E(i), A(j) \times B(k)}\\
&= \theta_{A \otimes B}(l = i \Concat j \Concat k)\;.
\end{align*}

\end{document}